\documentclass[aps,prc,onecolumn,showpacs,showkeys,nofootinbib,preprintnumbers]{revtex4-2}
\usepackage[paperwidth=21cm, paperheight=29.7cm, textwidth=17cm, textheight=24cm]{geometry}
\usepackage{graphicx}  
\usepackage{color}        
\usepackage{braket}
\usepackage{amsmath}
\usepackage{amssymb}
\usepackage{titlesec}
\usepackage{makecell}
\usepackage{natbib,hyperref}
\usepackage[mathscr]{euscript}
\usepackage{dutchcal}

\def \bea {\begin{eqnarray}}
\def \eea {\end{eqnarray}}

\usepackage{mathrsfs}  
\usepackage{bm}

\definecolor{ao-english}{rgb}{0.0, 0.5, 0.0}
\definecolor{cadmiumblue}{rgb}{0.0, 0.42, 0.24}

\begin{document}
	
\preprint{ECTP-2026-23}
\preprint{WLCAPP-2026-23}

\title{Chiral Phase Structure and In-Medium Modifications of Charmed Meson Masses within SU(4) extended Linear-Sigma Models}
\email{Corresponding author: atawfik@acu.edu.eg; 400778@iu.edu.sa; atawfik@bnl.gov}
	
\author{Nourhan~M.~Rfeek$^{1}$, Alexandra~Friesen$^{2}$, Yuri~Kalinovsky$^{2}$, Saleh~O.~Allehabi$^{3}$, Azzah~A.~Alshehri$^{4}$, Ashraf~F.~El-Sherif$^{5}$,  Abdel~Nasser~Tawfik$^{3,5}$}
	
\affiliation{$^{1}$Physics Department, Faculty of Science, Assiut University, 71515 Assiut, Egypt}
\affiliation{$^{2}$Joint Institute for Nuclear Research (JINR), 141980 Dubna, Russia Federation}
\affiliation{$^{3}$Department of Physics, Faculty of Science, Islamic University of Medinah (IU), Madinah 42351, Saudi Arabia} 
\affiliation{$^{4}$Department of Science and Technology, University College at Nairiyah, University of Hafr Al Batin (UHB), Nairiyah 31981, Saudi Arabia} 
\affiliation{$^{5}$Basic Science Department, Faculty of Engineering, Ahram Canadian University (ACU), 12556 Giza, Egypt}

\begin{abstract}
We extend the Polyakov-loop enhanced linear-sigma model to encompass four flavors and utilize it to analyze the light, strange, and charmed meson mass spectrum as it transitions from the vacuum into a hot and dense medium. Within the mean-field approximation, we determine the masses of scalar ($0^{++}$), pseudoscalar ($0^{-+}$), vector ($1^{--}$), and axial-vector ($1^{++}$) meson states, encompassing both open- and hidden-charm constituents, as functions of temperature $T$ and baryon chemical potential $\mu_B$. With parameters calibrated to the vacuum spectrum, the charm degree of freedom proves essential for an accurate description of observed masses, and charmed states acquire their in-medium modifications from the light and strange condensates that modify their valence quarks. Each species seems to undergo melting along its distinct trajectory, yet all transitions occur within a narrow band surrounding the chiral transition. The open-charm $D$-meson functions as a precise indicator of chiral restoration: its mass increases substantially across the transition and reflects the non-strange condensate, whereas hidden-charm states remain largely unaffected and serve as reference standards for the surrounding matter. We establish that transition points from three independent measurements delineate the chiral phase boundary ($T_c$, $\mu_B$), whose curvature matches lattice-QCD predictions and which persists as a crossover throughout the examined range, exhibiting no critical endpoint.
\end{abstract}.
	
\maketitle
	
	
\section{Introduction}

In order to investigate non-perturbative phenomena in Quantum Chromodynamics (QCD) with effective low-energy field theories, one needs to be able to study hadronic phenomena at finite temperatures, non-zero net-baryon densities and strong external electromagnetic backgrounds \cite{1,4}. Assuming massless quarks, the QCD Lagrangian maintains exact global chiral symmetry \cite{5,8}. The mechanism of dynamic chiral symmetry (DCS) breaking presumably produces condensate characterizing non-zero light-quark mass ($u$ and $d$). The explicit symmetry breaking in masses of light quarks leads to construction of the pseudo-Goldstone bosons (pions) having finite masses \cite{9,11}. In the case of heavier flavours ($s,c,\cdots$), this explicit symmetry breaking is further amplified leading to open-charm states such as $D$ mesons \cite{12} and hidden-charm states such as $\eta_c$ \cite{13}. At extreme temperature and/or baryon chemical potential (density), the vacuum chiral condensate expectedly melt and chiral symmetry will be restored with mass degeneracies among parity partners \cite{DeTar1989,Rapp2000}.
	
There are several theoretical approaches to explore the QCD phase diagram: statistical thermal approaches, such as the Hadron Resonance Gas (HRG) model \cite{20,21}, effective field theories, like the Nambu--Jona-Lasinio (NJL) model \cite{22,23},\ and the Linear-Sigma Model (LSM) \cite{3,24,25}, functional methods, such as Dyson--Schwinger equations \cite{16,18}, and Chiral Perturbation Theory \cite{19}. By fixing the LSM formulation to SU(2) flavor symmetry, pion dynamics can be reduced to two light flavors, providing a simpler framework to examine light-quark condensation \cite{24,26}. This can be extended to SU(3) where strange quark condensates can be included \cite{18} as well as mesonic nonets can be constructed \cite{3,23} and a deeper analysis of the QCD phase transition can be performed \cite{3,31}. Charm quarks can also be included under an SU(4) symmetric scheme and the dynamics of the charmed mesons can be analyzed, as well as the behavior of the charm-quark condensate. Using the closed-form analytic mass equations for charmed meson states, we calculate the thermal response of SU(4) implementations, thus extending to systematic studies of the in-medium modifications \cite{ref48,Eshraim2015}.
	
Such research endeavors are notably supported by the experimental undertakings carried out at relativistic heavy-ion collider installations such as CERN's Large Hadron Collider (LHC) and BNL's Relativistic Heavy Ion Collider (RHIC), especially with the STAR Beam Energy Scan. The shift of in-medium hadronic properties at high net-baryon densities will be further investigated in the future experimental projects at FAIR (GSI) \cite{37,39} and NICA (JINR) \cite{40}. In relativistic heavy-ion collisions, the severe environment defined by high temperatures and net-baryon densities generates a brief state of intensely interacting matter. Charmed hadrons, distinguished by their distinctive quark structure and relatively extended lifetimes, function as exceptionally valuable indicators of this medium. Their formation, reduction, and cooperative dynamics deliver vital information concerning deconfinement phenomena, quark-gluon plasma attributes, and hadronization mechanisms under such extreme conditions. Accordingly, the examination of charm-bearing hadronic particles constitutes an effective investigative methodology for discerning the fundamental architecture and temporal progression of intensely interacting matter produced in these collisions.

The modeling of the charm sector in effective theories such as the extended LSM (eLSM) is theory challenging because of the complicated coupling between light and heavy condensates, and explicit symmetry-breaking terms like $-2\,\mathrm{Tr}[\epsilon\Phi^\dagger\Phi]$~\cite{41}. Thermal properties of meson spectra in SU(3) have been successfully reproduced in the observables of light mesons. To overcome this, an extensive study is provided which examines the temperature in addition baryon chemical potential variation of scalar, pseudoscalar, vector, and axial-vector meson masses in the context of the SU(4) eLSM model. The thermodynamic potential includes the Polyakov-loop dynamics as well as the thermal excitations of quark-antiquark pairs in mean-field approximation. Moreover, pseudo-critical temperature values corresponding to chiral restoration phenomena will be determined, and the impact of the $U(1)_A$ axial anomaly coupling constant $c$ will be analyzed for different flavor scenarios.
	
As introduced, this article is devoted to the study of the in-medium modifications of meson masses for the full spectrum of light, strange, and charmed quark contents within the SU(4) eLSM framework. These masses are evaluated as functions of $T$ and $\mu_B$. Such an extensive analysis of both thermodynamic quantities is innovative and, to the authors' best knowledge, has not been previously explored. Prior investigations utilizing eLSM have been analyzing the spectra of light and strange quarks at finite  temperatures~\cite{31,Eshraim2015,SU34TempPRC} and baryon chemical potentials~\cite{TawfikSU4,Kovacs2016CEP,Kovacs2017PD}. To date, the masses of charmed meson states have been determined only in relation to temperatures~\cite{Eshraim2015,ms99-5qnq}. The in-medium modifications to light meson states driven by density (baryon chemical potential) seem to remain within the three-flavor framework~\cite{NuclearMatterPRC}. 
	
This paper is organized as follows: Section~\ref{sec:eLSM} summarizes the formulation of the extended Linear-Sigma Model in SU(4) schemes. In Section~\ref{secGrndPt}, the formalism of finite-temperature Polyakov-loop and construction of effective potential are presented; the grand canonical potential. The numerical results, the parameter fits and the evolution of the mass spectra are discussed in Section~\ref{sec:discussion}. The phase boundary are mapped out in Section~\ref{sec:phaseboundary}. Section~\ref{sec:conclusion} presents our concluding remarks and outlook.

\section{Extended Linear-Sigma Model}
\label{sec:eLSM}
The LSM provides a robust framework for modeling low-energy meson phenomenology \cite{ref49,ref54}. Expanding the flavor symmetry degree of freedom $N_f$ naturally broadens the spectrum of accessible mesonic states \cite{24,25,32}. Meson state representations in SU(3) were derived in refs.\cite{24,25}, whereas the extension to SU(4) was formulated in refs. \cite{32,ref55}. The total mesonic Lagrangian, which includes scalar, pseudoscalar, vector, and axial-vector channels, their interactions and the axial anomalies is presented as \cite{32}
\begin{equation}
\mathcal{L} = \mathcal{L}_{\text{SP}} + \mathcal{L}_{\text{VA}} + \mathcal{L}_{\text{Int}} + \mathcal{L}_{U(1)_A}.
\end{equation} 
In the case of the scalar-pseudoscalar kinetic and self-interaction terms, that are given by
\begin{equation}
\mathcal{L}_{\text{SP}} = \text{Tr}\left[(D^\mu \Phi)^\dagger (D_\mu \Phi)\right] - m^2 \text{Tr}(\Phi^\dagger \Phi) - \lambda_1 \left[\text{Tr}(\Phi^\dagger \Phi)\right]^2 - \lambda_2 \text{Tr}\left[(\Phi^\dagger \Phi)^2\right] + \text{Tr}\left[H(\Phi + \Phi^\dagger)\right].
\end{equation}
The vector and Axial-vector Lagrangian is expressed as
\begin{eqnarray}
\mathcal{L}_{\text{VA}} & = & -\frac{1}{4}\text{Tr}(L_{\mu\nu}^2 + R_{\mu\nu}^2) + \text{Tr}\left[\left(\frac{m_1^2}{2} + \Delta\right)(L_\mu^2 + R_\mu^2)\right] \\
& & +  \frac{g_2}{2}\left\{\text{Tr}(L_{\mu\nu}[L^\mu, L^\nu]) + \text{Tr}(R_{\mu\nu}[R^\mu, R^\nu])\right\} + g_3 \left[\text{Tr}(L_{\mu\nu} L^\mu L^\nu) + \text{Tr}(R_{\mu\nu} R^\mu R^\nu)\right] \\
&  & + g_4 \left[\text{Tr}(L_{\mu\nu} L^\nu L^\mu) + \text{Tr}(R_{\mu\nu} R^\nu R^\mu)\right] + g_5 \text{Tr}(L_{\mu\nu} R^{\mu\nu}) \\
&  & + g_6 \left[\text{Tr}(L_\mu L^\mu)\text{Tr}(L_\nu L^\nu) + \text{Tr}(R_\mu R^\mu)\text{Tr}(R_\nu R^\nu)\right].
\end{eqnarray}
The (axial) vector scalar mixing Lagrangian reads
\begin{equation}
\mathcal{L}_{\text{Int}} = \frac{h_1}{2}\text{Tr}(\Phi^\dagger \Phi)\text{Tr}(L_\mu^2 + R_\mu^2) + h_2 \text{Tr}(|L_\mu \Phi|^2 + |\Phi R_\mu|^2) + 2h_3 \text{Tr}(L_\mu \Phi R^\mu \Phi^\dagger).
\end{equation}
The axial $U(1)_A$ anomaly Lagrangian is given by~\cite{30}
\begin{equation}
\mathcal{L}_{U(1)_A} = c\left[\det(\Phi) + \det(\Phi^\dagger)\right] + c_0\left[\det(\Phi) - \det(\Phi^\dagger)\right]^2 + c_1\left[\det(\Phi) + \det(\Phi^\dagger)\right]\text{Tr}[\Phi\Phi^\dagger].
\end{equation}
The matrix fields representing scalar ($\sigma_a, J^{PC}=0^{++}$), pseudoscalar ($\pi_a, J^{PC}=0^{-+}$), vector ($V_a^\mu, J^{PC}=1^{--}$), and axial-vector ($A_a^\mu, J^{PC}=1^{++}$) meson multiplets are expanded as \cite{29}
\begin{equation}
\Phi = \sum_{a=0}^{N_f^2-1} T_a (\sigma_a + i\pi_a), \quad L^\mu = \sum_{a=0}^{N_f^2-1} T_a (V_a^\mu + A_a^\mu), \quad R^\mu = \sum_{a=0}^{N_f^2-1} T_a (V_a^\mu - A_a^\mu) .
\end{equation}
In this case $T_a = \hat{\lambda}_a/2$ are the generators of the flavor group $U(N_f)$, where the $\hat{\lambda}_a$ are the generalized Gell-Mann matrices. The coupling between the scalar and vector sectors in the gauge covariant derivative is provided by
\begin{equation}
D^\mu \Phi \equiv \partial^\mu \Phi - i g_1 (L^\mu \Phi - \Phi R^\mu) - ie A^\mu [T_3, \Phi],
\end{equation}
where the mesonic gauge coupling $g_1$ is used. The tensors of the left- and right-handed vector fields are expressed as \cite{29}
\begin{eqnarray}
L^{\mu\nu} &\equiv & \partial^\mu L^\nu - ie A^\mu [T_3, L^\nu] - (\partial^\nu L^\mu - ie A^\nu [T_3, L^\mu]), \\
R^{\mu\nu} &\equiv & \partial^\mu R^\nu - ie A^\mu [T_3, R^\nu] - (\partial^\nu R^\mu - ie A^\nu [T_3, R^\mu]),
\end{eqnarray}
where $A^{\mu}=g A_\mu^a \lambda^a/2$ is the external field. The constant $g$ is the Yukawa coupling, which is fixed from the non-strange constituent quark mass. The main chiral structure is preserved when the flavor symmetry is extended from $N_f=3$ to $N_f=4$ (except for the presence of an explicit symmetry-breaking mass term $\mathcal{L}_{\text{emass}} = - 2 \text{Tr}[\epsilon\Phi^\dagger\Phi]$ which is necessary to explain the large charm mass scale). 
	
For the tree-level $SU(4)$ mesonic potential, in the rotated non-strange ($\sigma_x$), strange ($\sigma_y$), and charm ($\sigma_c$) condensate basis, the potential reads
\begin{eqnarray}
U(\sigma_x, \sigma_y, \sigma_c) & = & \frac{1}{2}m^2(\sigma_x^2 + \sigma_y^2 + \sigma_c^2) - \frac{c}{4}\sigma_x^2 \sigma_y \sigma_c + \frac{\lambda_1}{2}(\sigma_x^2 \sigma_y^2 + \sigma_x^2 \sigma_c^2 + \sigma_y^2 \sigma_c^2) \\
& + & \frac{1}{8}(2\lambda_1 + \lambda_2)\sigma_x^4 + \frac{1}{4}(\lambda_1 + \lambda_2)\sigma_y^4 + \frac{1}{4}(\lambda_1 + \lambda_2)\sigma_c^4 \\
& - & h_x \sigma_x - h_y \sigma_y - h_c \sigma_c + \epsilon_c \sigma_c^2.
\end{eqnarray}
Minimization with respect to the vacuum fields gives the following conditions for field alignment of $h_x, h_y$, and $h_c$, so that
\begin{align}
h_x &= m^2\sigma_x - \frac{c}{2}\sigma_x \sigma_y \sigma_c + \lambda_1\sigma_x(\sigma_y^2 + \sigma_c^2) + \frac{1}{2}(2\lambda_1 + \lambda_2)\sigma_x^3, \\
h_y &= m^2\sigma_y - \frac{c}{4}\sigma_x^2\sigma_c + \lambda_1\sigma_y(\sigma_x^2 + \sigma_c^2) + (\lambda_1 + \lambda_2)\sigma_y^3, \\
h_c &= m^2\sigma_c + 2\epsilon_c\sigma_c - \frac{c}{4}\sigma_x^2\sigma_y + \lambda_1\sigma_c(\sigma_x^2 + \sigma_y^2) + (\lambda_1 + \lambda_2)\sigma_c^3.
\end{align}
This work builds upon our $SU(4)$ configuration in ref.~\cite{ref48}; the parameter values, except for $h_1$, and the tree-level meson masses follow from the quadratic terms of the Lagrangian and are detailed there.

\section{The Grand Canonical Potential}
\label{secGrndPt}

We utilize the Polyakov-loop extended LSM (PLSM) \cite{1,24} to take into account the thermal and dense medium effects. The system Lagrangian is a sum of chiral mesonic sector and Polyakov-loop potential \cite{1,24}
\begin{equation}
\mathcal{L} = \mathcal{L}_{\text{mes}} + \mathcal{L}_{\bar{q}q} - \mathcal{U}(\phi, \bar{\phi}, T), 
\end{equation}
with $\mathcal{L}_{\bar{q}q} = \sum_f\bar{q_f}(i\gamma^\mu D^\mu - g T_a(\sigma_a+ i\gamma_\mu \pi_a))q_f$. The path integral expression of the grand canonical partition function in the thermal equilibrium is derived as
\begin{equation}
\mathcal{Z} = \text{Tr}\exp(-\hat{\mathcal{H}}/T) = \int \prod_a \mathcal{D}\sigma_a \int \mathcal{D}q \mathcal{D}\bar{q} \exp\left[\int_x \mathcal{L} + \sum_{f}\mu_f\bar{q}\gamma^0q\right],
\end{equation}
where $\int_x \equiv i \int_0^{1/T} dt \int_V d^3x$ and $\sigma_a$ sums up all boson fields. In the mean-field approximation, this is done by replacing the quantum meson fields with the thermal averages of the meson field operators \cite{32} and the chemical potential contains all vector fields $\mu_f = \mu_f - g_{f\omega}\omega - g_{f \rho} t^3 \rho -  g_{f \phi} \phi - g_{f J/\Psi} \psi$. 

In this work, we consider isospin-symmetric quark-meson matter without explicitly including baryonic degrees-of-freedom and can neglect all vector interactions in the quark sector. In this regime, the $\rho$-field does not contribute due to isospin symmetry, the heavy mesons are suppressed by their large masses at the considered scale, and the contribution of the $\omega$-meson to the effective quark chemical potentials is negligibly small, since the mean field $\omega$ is not generated in the absence of a baryonic current \cite{Dexheimer:2008}. The grand potential may be reformulated as
\begin{equation}
\Omega(T,\mu_B) = \frac{-T\ln\mathcal{Z}}{V} = U(\sigma_x,\sigma_y,\sigma_c)+\mathcal{U}(\phi,\bar\phi,T) + \Omega_{q\bar q}(T,\mu_B),
\end{equation}
and seems to lack all vector condensates; vector couplings emerge only within tree-level meson mass formulations. Our equation of state thus corresponds to the $g_{\omega}=g_{\rho}=0$ limit, featuring no density-dependent vector repulsive effects. We present this characterization in precisely this manner rather than claim to reproduce the claims of ref. \cite{Dexheimer:2008}, which represents a chiral mean-field model that \emph{contains} the $\omega,\rho,\phi$ fields: we share merely its vanishing-vector limit.
	
In this potential, the complex Polyakov-loop order parameters $\phi$ and $\bar{\phi}$ and Polyakov-loop potential $\mathcal{U}(\phi, \bar{\phi}, T)$ is used in the logarithmic form \cite{RRW}. The vacuum potential $U(\bar{\sigma})$ corresponds to the tree-level mesonic interactions and ground-state properties in vacuum \cite{32}. Quark-antiquark thermal potential $\Omega_{q\bar{q}}$ includes thermal Fermi-gas excitation and constituent quark mass modifications \cite{ref54}
\begin{equation}
\Omega_{q\bar{q}} = - 2 \nu_c T \sum_f \int \frac{d^3p}{(2\pi)^3} E_f - 2 \nu_c T \sum_f \int \frac{d^3p}{(2\pi)^3} (\ln g_f^+ + \ln g_f^-),
\end{equation}
where $\nu_c=2N_c=6$ is the color-spin degeneracy, $f$ sums over the active quark flavors ($u, d, s, c$). The constituent masses $m_{u,d}=g\sigma_x/2$, $m_s=g\sigma_y/\sqrt2$, $m_c=g\sigma_c/\sqrt2$. The Yukawa coupling $g=2m_q/\bar{\sigma}_x$ is the ratio of the non-strange constituent quark mass $m_q$ to the condensate $\sigma_x$.  Thermal distribution functions $g_f^+$ and $g_f^-$~\cite{24} are defined as
\begin{eqnarray}
g_f^{+} &=& 1 + 3(\phi + \bar{\phi} e^{-(E_f - \mu_f)/T}) e^{-(E_f - \mu_f)/T} + e^{-3(E_f - \mu_f)/T} \label{eq:gfp} \\
g_f^{-} &=& 1 + 3(\bar{\phi} + \phi e^{-(E_f + \mu_f)/T}) e^{-(E_f + \mu_f)/T} + e^{-3(E_f + \mu_f)/T}. \label{eq:gfm}
\end{eqnarray}
The order parameters $(\sigma_x, \sigma_y, \sigma_c, \phi, \bar{\phi})$ at a particular temperature $T$ and baryon chemical potential $\mu_B$ are determined from the set of stationary conditions for $\text{Re}(\Omega)$~\cite{32}.

We proceed to expound upon additional particulars concerning the chemical potential. The analysis presumes a singular conserved charge -- baryon number -- and consequently a single chemical potential $\mu_B$. Given that each quark possesses a baryon number of $1/3$, this identical potential manifests across all flavour types,
\begin{equation}
\mu_u=\mu_d=\mu_s=\mu_c=\frac{\mu_B}{3}.
\end{equation}
The strange and charm quarks consequently do not receive distinct chemical potentials; the independent isospin, strangeness, and charm potentials are set to zero, namely $\mu_I=\mu_S=\mu_C=0$. Each quark flavor experiences solely $\mu_f=\mu_B/3$ via the distribution functions, Eqs.~\eqref{eq:gfp}--\eqref{eq:gfp}.
	
When $\mu_B=0$, the conditions $\mu_f=0$ and $\phi=\bar\phi$ are satisfied, which results in $g_f^{+}=g_f^{-}$. Under these circumstances, the contributions from quarks and antiquarks precisely offset one another, causing all net flavour densities to equal zero exactly,
\begin{equation}
\langle n_s\rangle=\langle n_c\rangle=\langle n_l\rangle=0,
\qquad n_f=-\frac{\partial\Omega_f}{\partial\mu_f}.
\end{equation}
This precise neutrality is maintained along the temperature axis (with numerical values $|n_s|,|n_c|\lesssim10^{-12}\,\mathrm{GeV}^3$). The density-dependent findings -- specifically the mass surfaces within the $(T,\mu_B)$ plane and the chiral phase boundary - employ the identical specification $\mu_S=\mu_C=0$, maintained uniformly throughout the analysis. In this framework, the chemical potentials are constrained rather than the densities: a modest net strange density emerges (whereas the charm density remains negligible due to $m_c\gg\mu_B/3$), therefore we do not assert exact strangeness and charm neutrality at finite $\mu_B$, which would necessitate self-consistent resolution for $\mu_S,\mu_C\neq0$, representing an alternative (strangeness-neutral) framework. The finite-$\mu_B$ calculations are executed under the condition $\mu_S=\mu_C=0$ and remain valid for this specification. The reason for this choice is the mapping of the spectrum and the chiral boundary along the baryon axis, i.e., matter is not modeled with a fixed net strangeness or charm, so no additional chemical potential is warranted. It is the standard single-$\mu_B$ prescription of PLSM/PQM studies.

\section{Results and Discussion}
\label{sec:discussion}
	
\begin{figure}[htbp]
\centering
\includegraphics[width=0.48\linewidth]{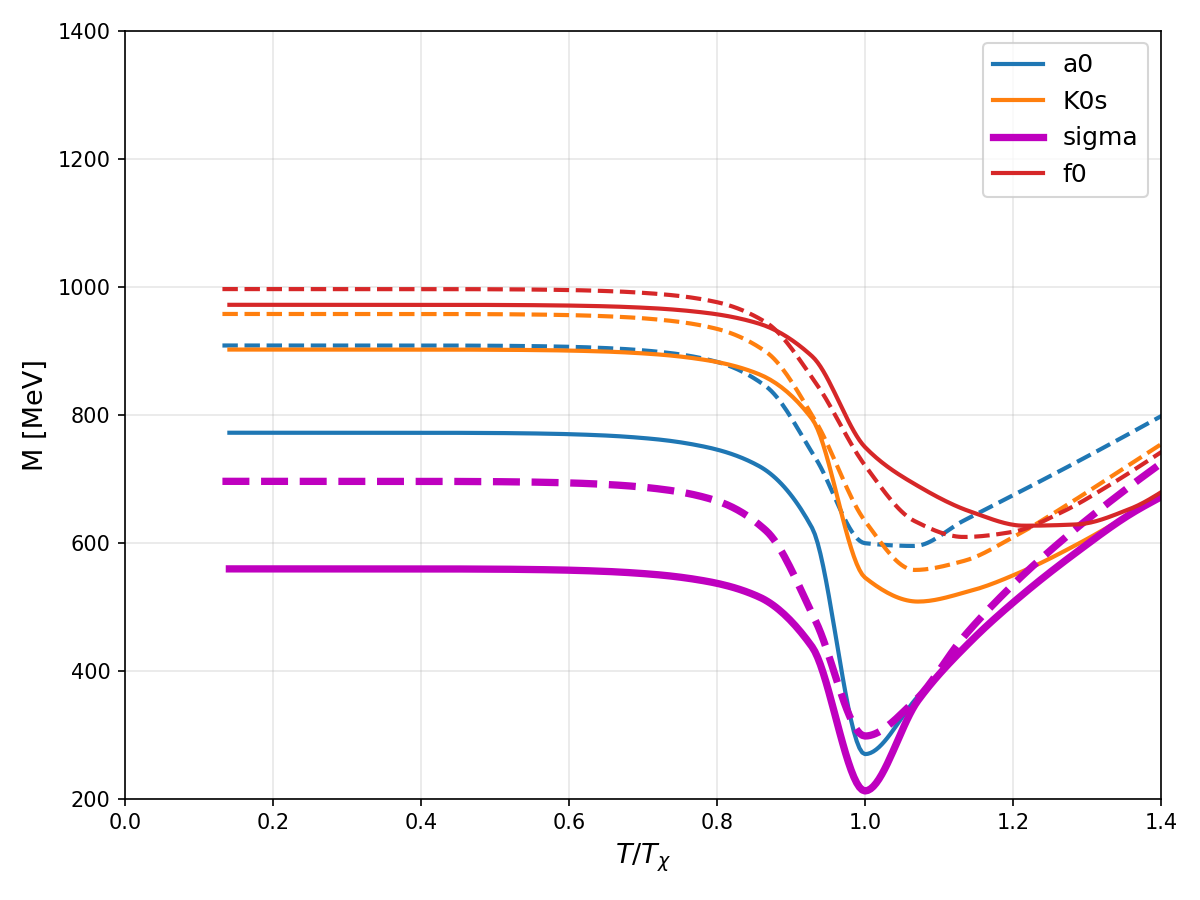}		\hfill
\includegraphics[width=0.48\linewidth]{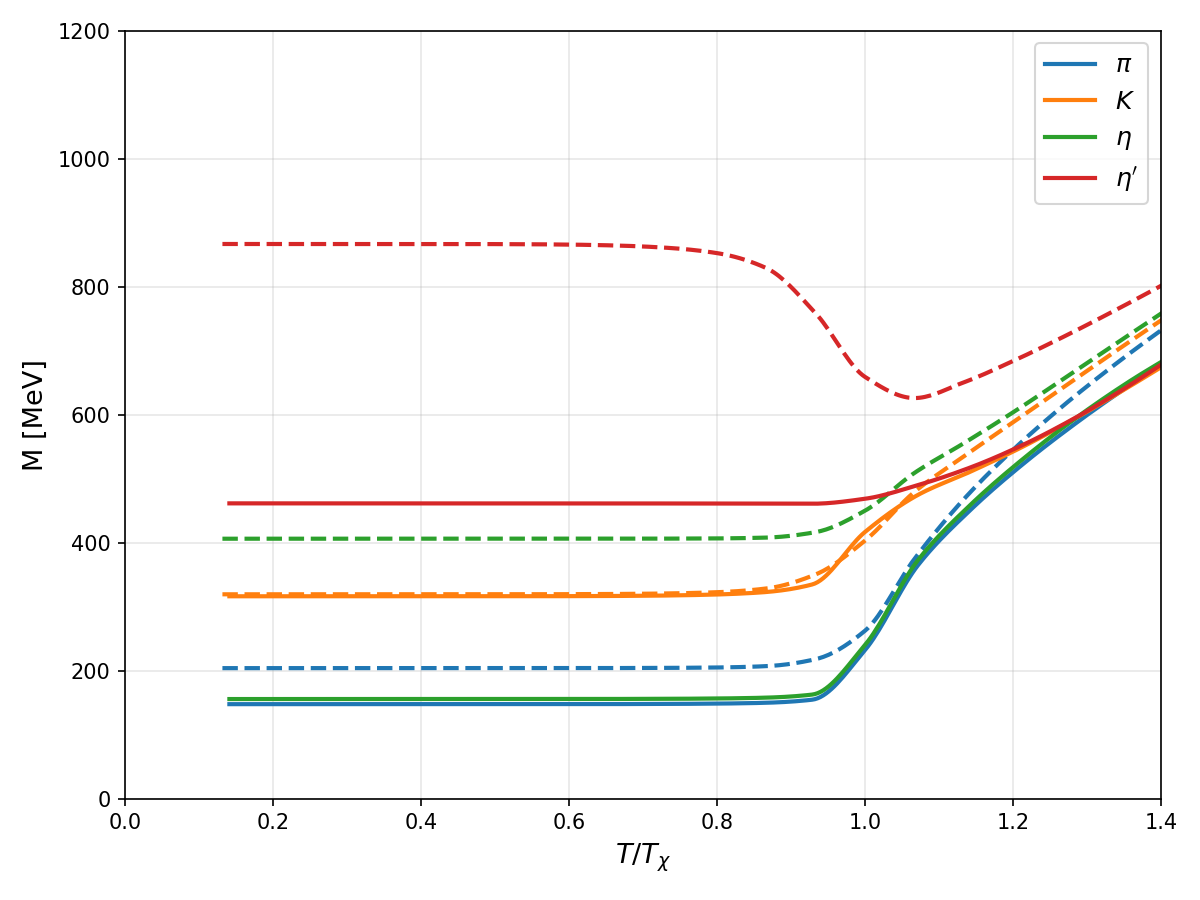}
\caption{The temperature dependence of scalar ($0^{++}$ left panel) and pseudoscalar ($0^{-+}$ right panel) meson masses as functions of temperature and vanishing baryon chemical potential for SU(4) at $c=0$ (solid) and $c=11.24$ (dashed line). }
\label{Fig:1}
\end{figure}

\begin{figure}[htbp]
\centering
\includegraphics[width=0.5\linewidth]{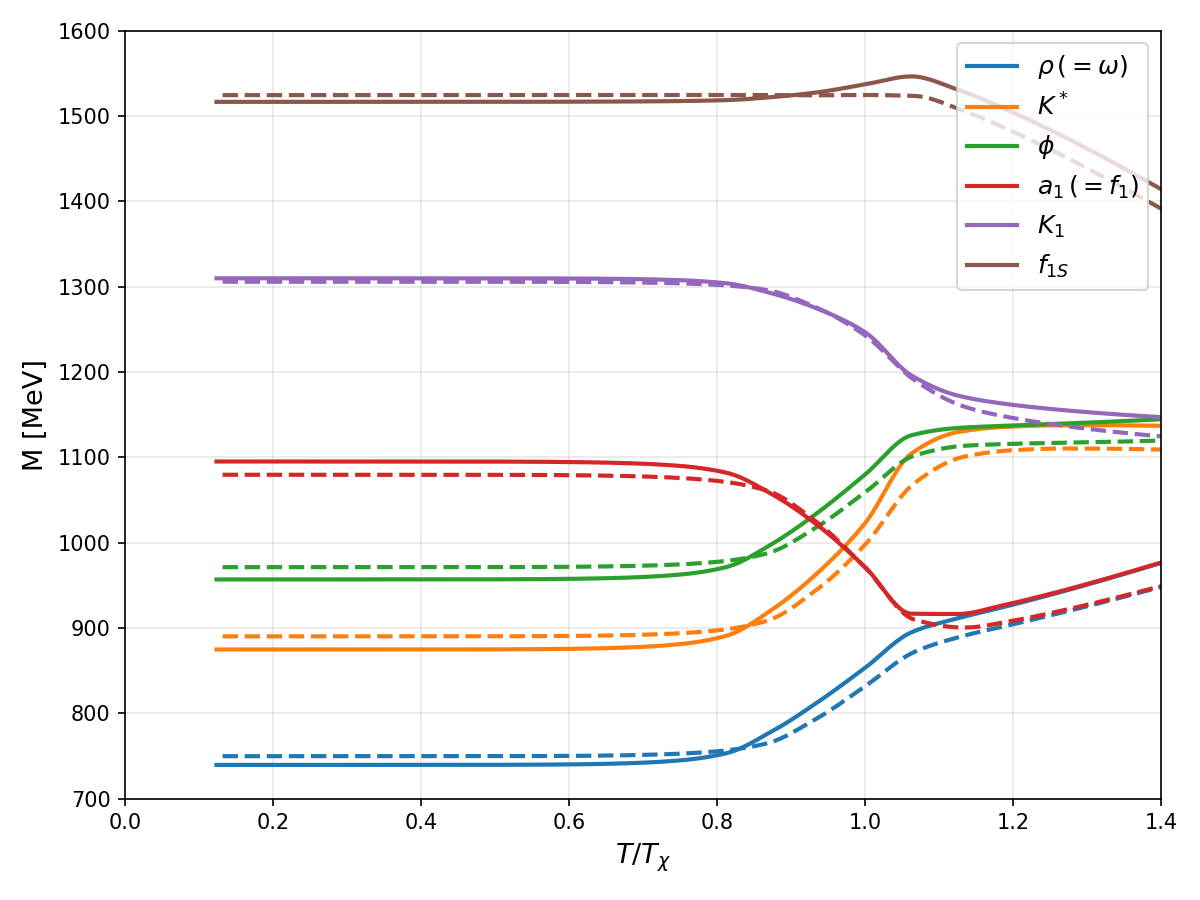}
\caption{The same as in Fig.~\ref{Fig:1} but here vector and axial-vector mesons.}
\label{Fig:2}
\end{figure}

%
We utilize the mass equations and the SU(4) parameter set from Table~II in the paper \cite{SU34TempPRC}. The masses depicted in Figs.~\ref{Fig:1}--\ref{Fig:2} are evaluated at $\mu_B=0$ and are plotted against the reduced temperature $T/T_{\chi}$, where $T_{\chi}$ corresponds to the pseudo-critical temperature of the chiral phase transition, $T_{\chi}=198$~MeV for $c=0$ and $T_{\chi}=208$~MeV for $c\neq0$  \cite{SU34TempPRC}. 
	
The left panel of Fig.~\ref{Fig:1} displays the scalar meson masses, whereas the right panel illustrates the pseudoscalar meson masses as functions of temperature. The behavior of mesons demonstrates a clear pattern: as temperature increases in the vicinity of and beyond the chiral phase transition ($T/T_{\chi}\approx 1$), pseudoscalar meson masses show minimal change at lower temperatures before experiencing a substantial increase in the high-temperature regime. Scalar mesons soften: $a_0, K^s_0, \sigma, f_0$ remain approximately constant at low temperatures, subsequently decrease as $T\rightarrow T_{\chi}$, reach a minimum near the critical temperature, and subsequently increase. Pseudoscalar mesons harden: the masses of $\pi, K, \eta, \eta'$ remain flat at low temperatures and approach their scalar chiral partners, indicating restored chiral symmetry. The $U(1)_A$ term influences only the isoscalar pseudoscalars, i.e., $c=0$, the $\eta'$ mass is nearly equivalent to that of the pion, whereas when $c\neq0$, it is substantially elevated above the octet.

%
The vector and axial-vector nonets are presented in Fig.~\ref{Fig:2}. The observed pattern mirrors that found in the scalar-pseudoscalar sector: axial-vector masses soften and diminish as temperature increases, while vector masses harden and increase with rising temperature. These two sets of masses converge toward a common value in the vicinity of $T_{\chi}$, resulting in parity-partner degeneracy across the phase transition. Given that the (axial-)vector masses receive negligible contribution from the $\rm{det} \Phi$ term, they demonstrate substantial insensitivity to the parameter $c$, causing the two sets of curves to nearly coincide.

Collectively, Figs.~\ref{Fig:1}--\ref{Fig:2} demonstrate that each meson state exhibits a distinct state-dependent trajectory, although dissociation transpires within a confined region proximate to $T_\chi$, demonstrating minimal divergence across different states. The direction of the mass shift is determined by the channel's polarity characteristics (scalars and axial-vectors undergo softening; pseudoscalars and vectors undergo hardening), irrespective of flavor considerations.

Prior to extensive examining the in-medium modifications of mason masses, we present the vacuum charmed masses determined in this investigation and provide a comparison with experimental data. As demonstrated in Tab.~\ref{tab:charmed_exp}, the results demonstrate substantial agreement with the experimental observations, with deviations remaining within a few percent~\cite{32,SU34TempPRC}: $\eta_c$ and $\eta_{c0}$ exhibit close correspondence with experimental values; however, $J/\Psi$ is notably lower than its measured value, and $\chi_{c1}$ surpasses anticipated values. This observation is consistent with a tree-level approach grounded in light-quark chiral symmetry~\cite{Parganlija2010,31,Eshraim2015,Eshraim2018}. It should be emphasized that none of the vacuum offsets affect the patterns examined in the subsequent sections. These patterns are governed by the behavior of the condensates $\sigma_x, \sigma_y, \sigma_c$, rather than by the relative magnitudes.

\begin{table}[htbp]
\centering
\begin{tabular}{c | c | c}
\hline\hline
Charmed Meson State & This Work [MeV] & Exp. Results [MeV] \\
\hline\hline
\multicolumn{3}{l}{Pseudoscalar $(0^{-+})$}\\
$D$        & 2147 & 1869 \\
$D_s$      & 2413 & 1968 \\
$\eta_c$   & 3016 & 2984 \\
\hline
\multicolumn{3}{l}{Scalar $(0^{++})$}\\
$D_0^{*}$    & 2735 & 2343 \\
$D_{s0}^{*}$ & 2612 & 2318 \\
$\chi_{c0}$  & 3504 & 3415 \\
\hline
\multicolumn{3}{l}{Vector $(1^{--})$}\\
$D^{*}$    & 2112 & 2010 \\
$D_s^{*}$  & 2053 & 2112 \\
$J/\Psi$   & 2352 & 3097 \\
\hline
\multicolumn{3}{l}{Axial-vector $(1^{++})$}\\
$D_1$      & 2628 & 2422 \\
$D_{s1}$   & 2814 & 2460 \\
$\chi_{c1}$ & 3933 & 3511 \\
\hline
\end{tabular}
\caption{The vacuum masses of the charmed mesons (in MeV) derived in this study using the SU(4) eLSM are categorized by spin-parity and contrasted with the existing experimental data~\cite{ref62}. }
\label{tab:charmed_exp}
\end{table}
	
The flavor hierarchy manifests itself distinctly at vanishing baryon chemical potential ($\mu_B=0$): strange hadronic states display a considerable postponement in their thermal response in comparison to their non-strange analogs. In contrast, quarkonium states remain exceptionally resilient, exhibiting negligible or minimal susceptibility to thermal variations. This lack of responsiveness provides an early signal of the Silver--Blaze phenomenon \cite{Cohen2003,Braun2021}, wherein particular observables remain unaffected by modifications in temperature or chemical potential until a critical boundary is surpassed -- a concept that will be further elaborated in forthcoming discussion. The continuation of this pattern at zero baryon chemical potential has been previously noted in ref.~\cite{ms99-5qnq}, underscoring its relevance as a fundamental attribute of strongly interacting matter.

\subsection{(T, $\mu_B$) Scan}

\begin{figure}[htbp]
\centering
\includegraphics[width=0.48\textwidth]{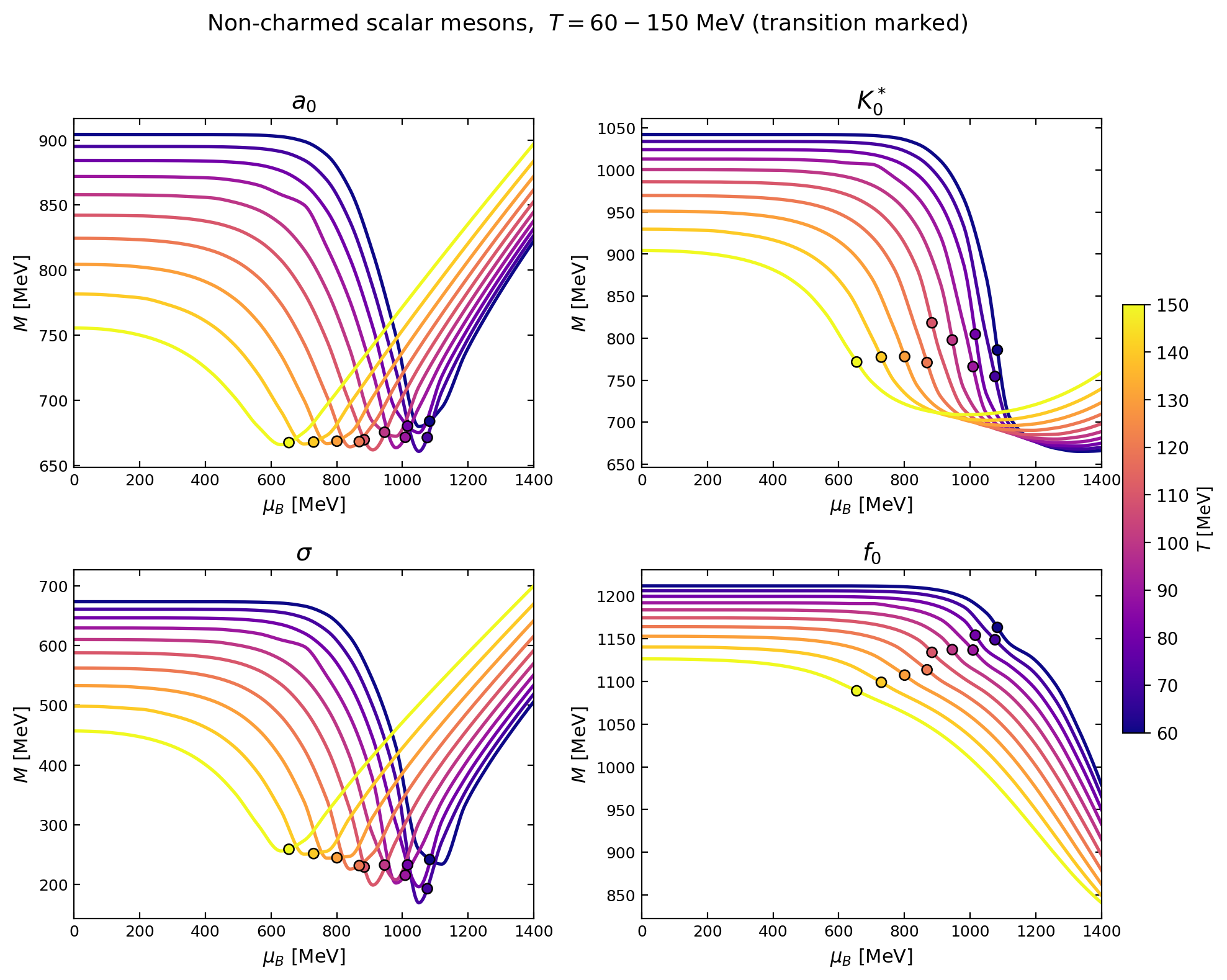} \hfill
\includegraphics[width=0.48\textwidth]{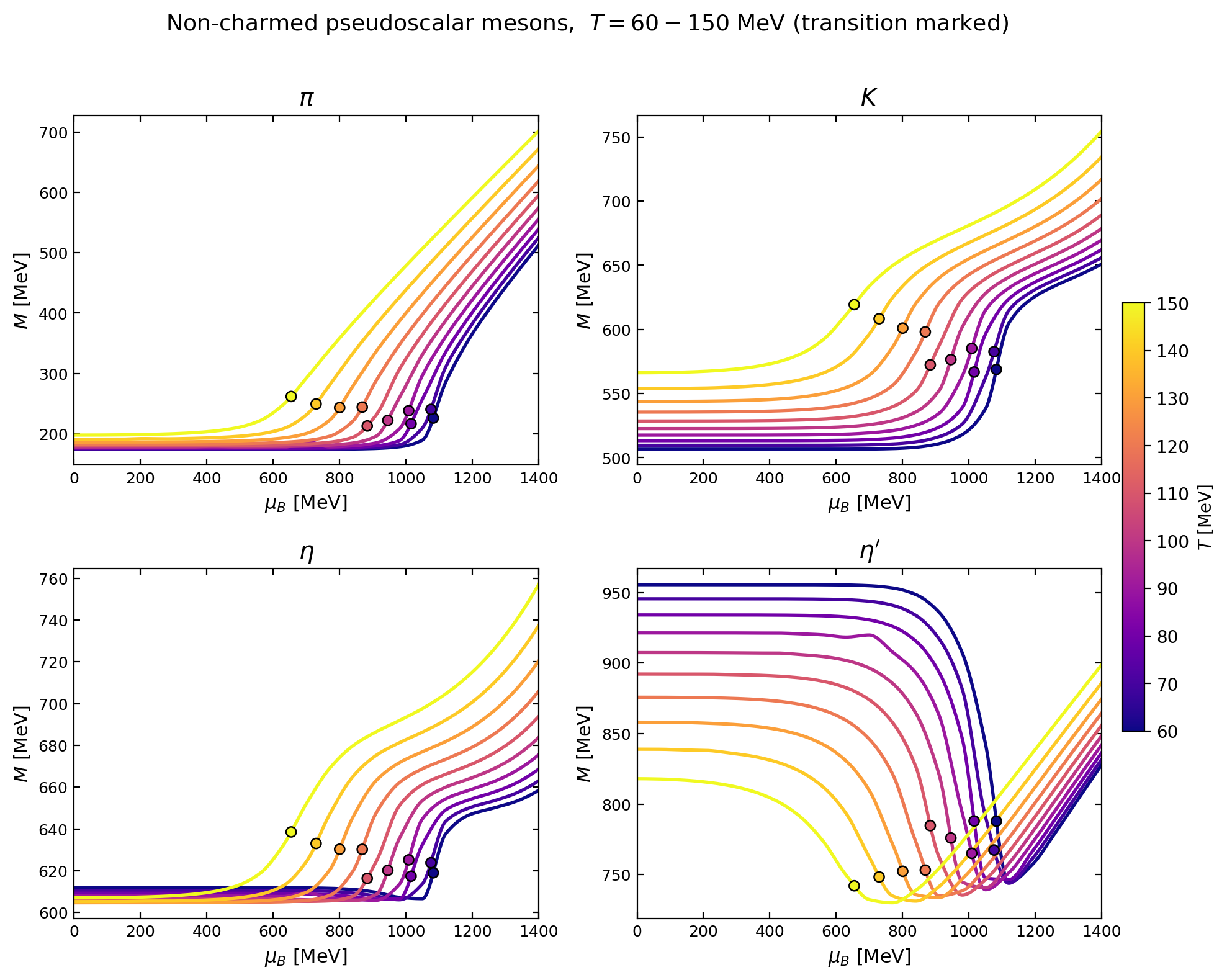} 
\includegraphics[width=0.48\textwidth]{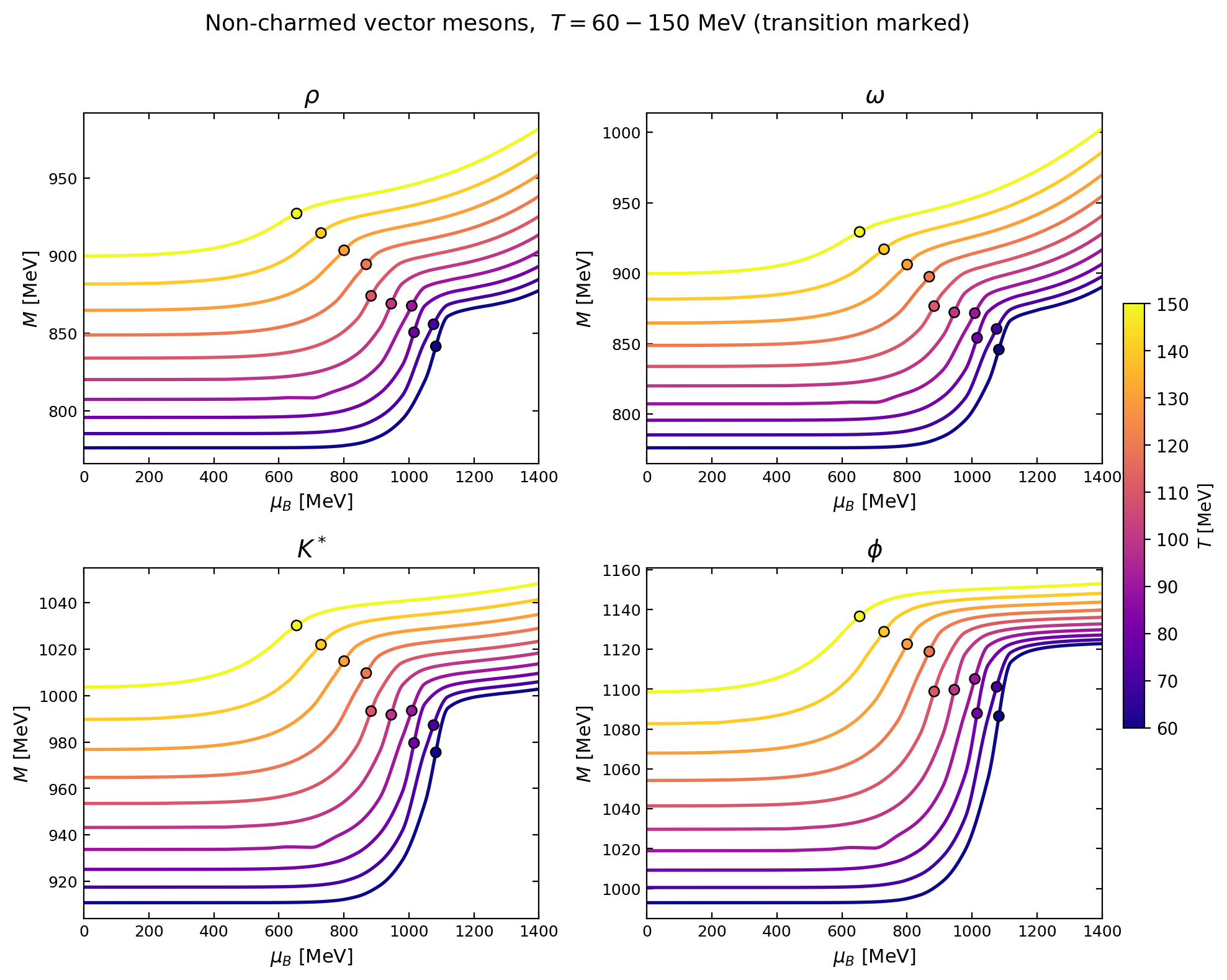} \hfill
\includegraphics[width=0.48\textwidth]{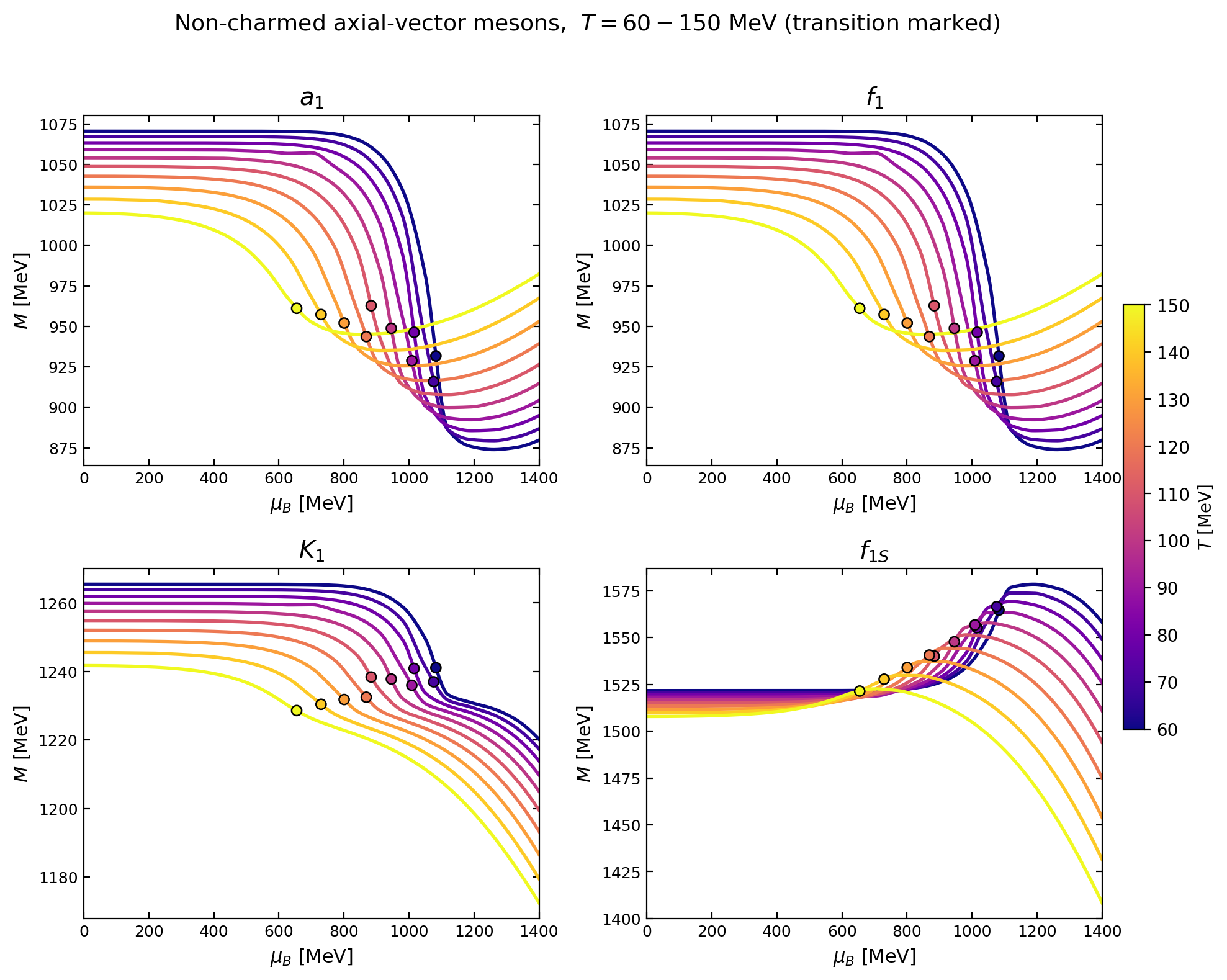}
\caption{The evolution of effective masses for non-charmed mesons across scalar ($a_0, K_0^*, \sigma, f_0$), pseudoscalar ($\pi, K, \eta, \eta'$), vector ($\rho, \omega, K^*, \phi$), and axial-vector ($a_1, f_1, K_1, f_{1s}$) as functions $\mu_B$ within the temperature range of $T=60$ to $150\text{ MeV}$. The plot demonstrates the manifestation of channel-dependent mass splitting phenomena, significant softening effects in the vicinity of the chiral phase transition, and alterations to meson properties occurring within hot and dense hadronic matter environments. The circular markers denote the respective chiral transition temperature. }
\label{Fig:nonCh_mu}
\end{figure}

\begin{figure}[htpt]
\centering
\includegraphics[width=0.85\textwidth]{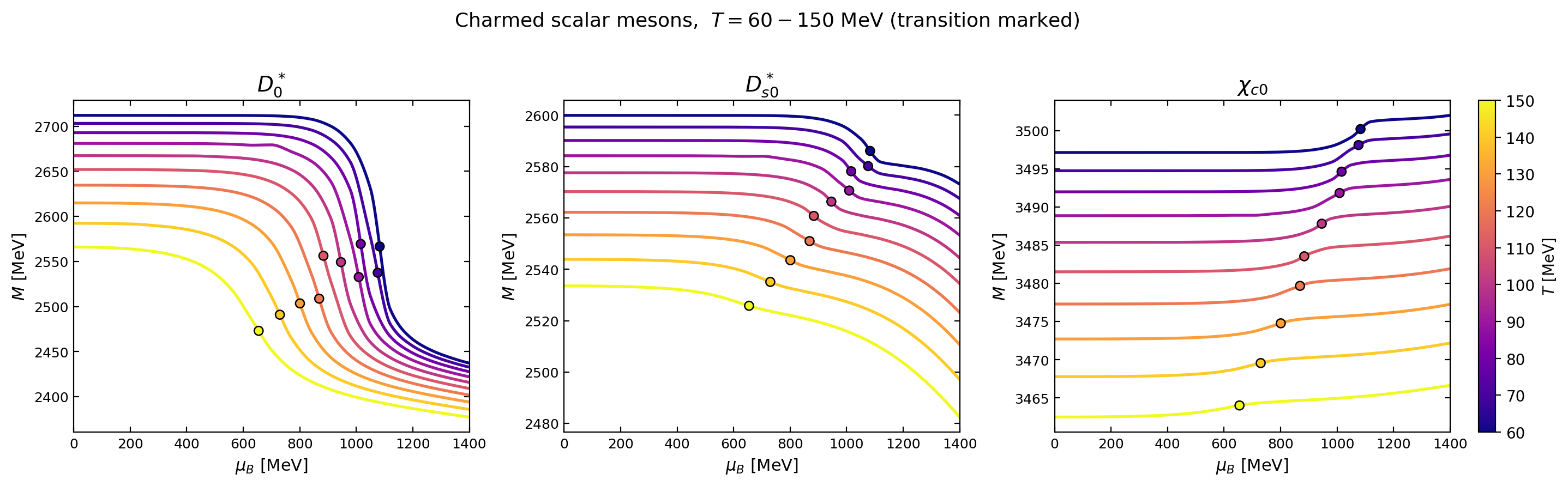}\hfill  
\includegraphics[width=0.85\textwidth]{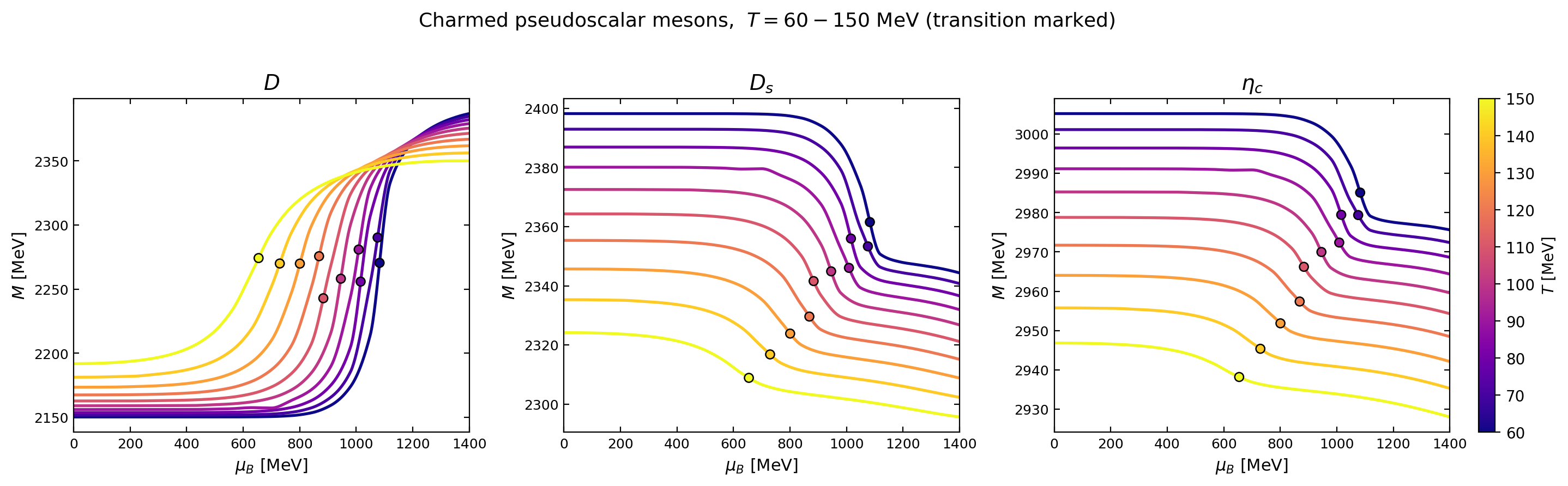}\hfill
\includegraphics[width=0.85\textwidth]{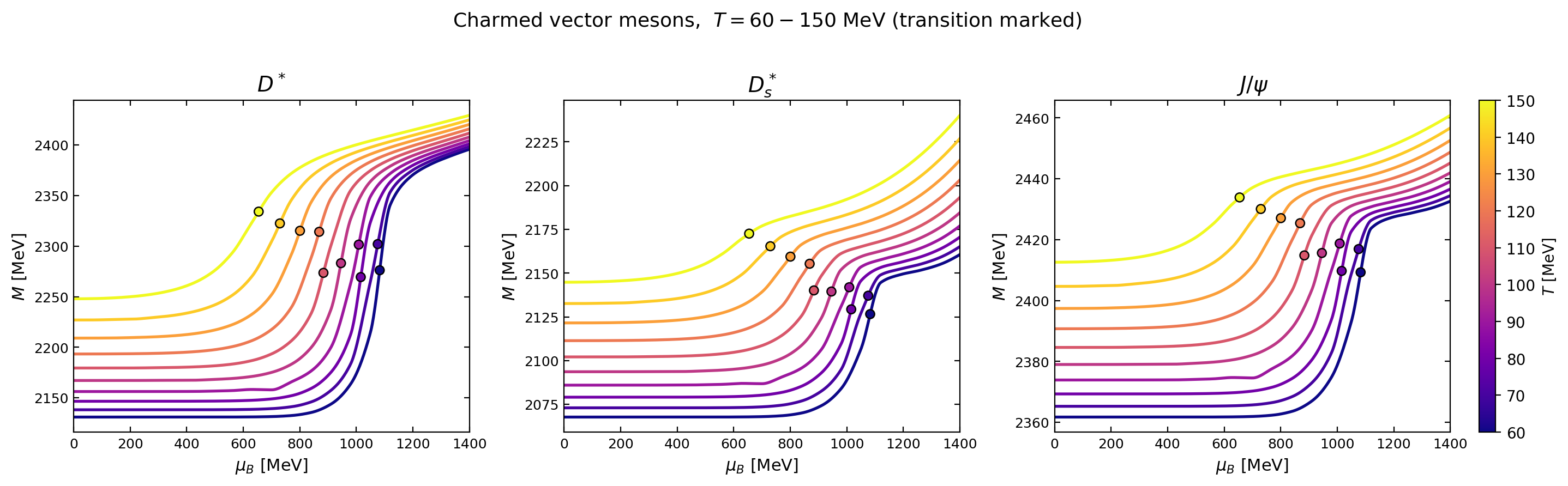}\hfill
\includegraphics[width=0.85\textwidth]{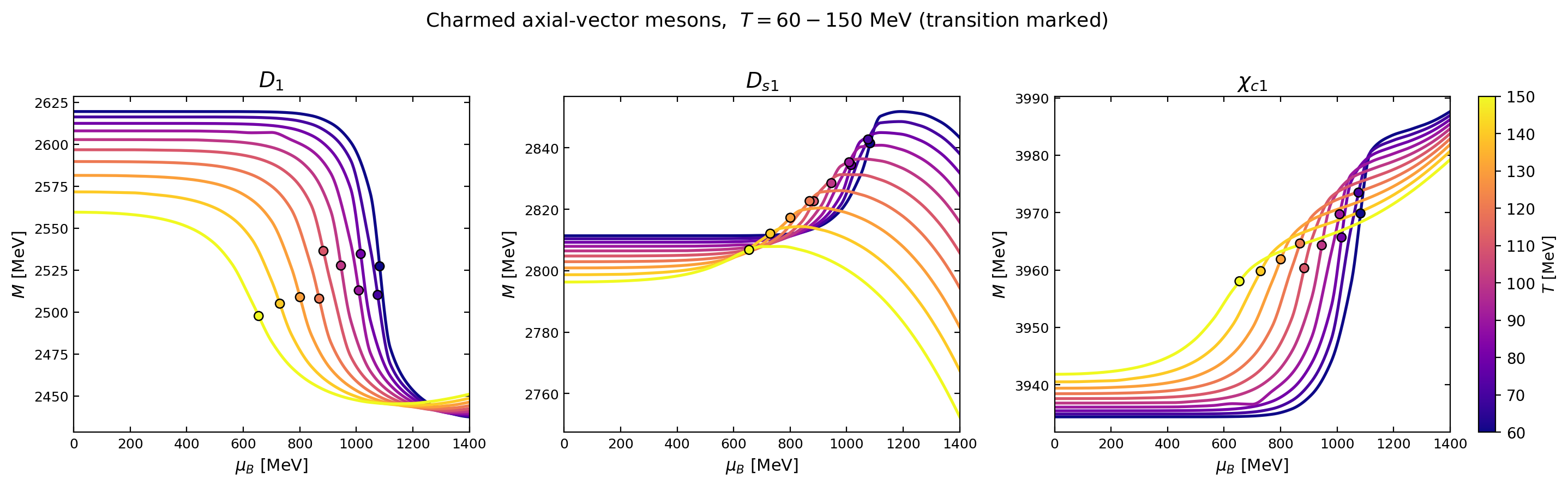}\hfill
\caption{The same as in Fig.~\ref{Fig:Ch_mu} but here for charmed meson states, including scalar ($D_0^*$, $D_{s0}^*$, $\chi_{c0}$), pseudoscalar ($D$, $D_s$, $\eta_c$), vector ($D^*$, $D_s^*$, $J/\Psi$), and axial-vector ($D_1$, $D_{s1}$, $\chi_{c1}$) meson states. }
\label{Fig:Ch_mu}
\end{figure}

%
The comprehensive spectrum is presented in Fig. \ref{Fig:nonCh_mu} (non-charmed nonets) and Fig. \ref{Fig:Ch_mu} (charmed sector), which demonstrate the behavior in which effective meson masses vary as functions of $\mu_B$ throughout an extensive thermal interval spanning from $T=60\text{ MeV}$ to $150\text{ MeV}$ at $c = 11.24$. The circular symbols designate the chiral temperature, the point at which (hadronic) meson states are dissolved into (partonic) quark states. To preserve clarity and maintain focus within the primary exposition, the complete collection of three-dimensional response surfaces providing detailed information on each individual meson state has been relegated to Appendix~\ref{app:lowT} (Figs. \ref{Fig:7}--\ref{Fig:10}), where all panels present mass spectra both for $c=0$ and $c=11.24$.

For scalar and axial-vector mesons, the vacuum expectation value exhibits a systematic decrease toward lower baryon chemical potential as temperature increases -- manifesting as a sharp transition near $\mu_B\approx 900--1000$~MeV at $T=100$~MeV and progressively softening into a gentle crossover by $T=300$~MeV - thereby establishing the phase boundary through the family of temperature-dependent curves. Within the non-charmed sector, vector states demonstrate hardening behavior beyond the onset point: the degenerate $\rho=\omega$ pair acquires approximately $80$~MeV at $\mu_B=1200$~MeV, $K^*$ increases from approximately $943$ to approximately $1015$~MeV, and the nearly pure $\bar{s}s$ state $\phi$ exhibits minimal movement, reaching approximately $1130$~MeV because its $\sigma_y$ dressing undergoes delayed melting. The axial-vector mesons display opposite behavior: the degenerate $a_1=f_1$ decreases from approximately $1054$ to approximately $900$~MeV and intersects the ascending $\rho=\omega$. $K_1$ undergoes mild downward drift from approximately $1258$~MeV, while the nearly pure-strange $f_{1s}$ remains near $1520$~MeV. In the scalar sector, every state softens throughout the transition -- $a_0$ from $858$ to $773$~MeV, $K^*_0$ from $1000$ to $681$~MeV, $f_0$ from $1184$ to $1049$~MeV -- with the $\sigma$ descending from $610$~MeV into a pronounced minimum representing the soft mode of the chiral transition. The pseudoscalar sector exhibits analogous behavior: $\pi, K, \eta$ remain flat on the plateau before subsequently increasing as they forfeit their Goldstone protection, whereas the highest state, $\eta'$, instead decreases from approximately $907$~MeV as the $U(1)_A$ anomaly contribution diminishes. Beyond the transition, each pseudoscalar converges toward its scalar counterpart - $\pi$ with $\sigma, K$ with $K^*_0, \eta'$ with $f_0$ -- providing direct spectral evidence of chiral-symmetry restoration. Within each family, strange-containing members lag their non-strange counterparts, as $\sigma_y$ persists to larger baryon chemical potential values than $\sigma_x$.

The charmed spectrum is classified into the same four families and is differentiated based on whether a state contains a light valence quark. In the charmed-pseudoscalar panel, $D(c\bar{q})$ increases beyond onset, $D_s (c\bar{s})$ exhibits minimal change, and $\eta_c (c\bar{c})$ remains constant; the charmed-vector panel demonstrates identical behavior, with $D^*$ increasing by approximately $200$~MeV between the plateau and $\mu_B=1200$~MeV, $D_s^*$ increasing marginally, and $J/\Psi$ remaining essentially unchanged. The charmed-scalar and charmed-axial-vector panels exhibit opposite trends: $D_0^*$ and $D_1$ decline gradually beyond onset, paralleling their light scalar and axial counterparts, whereas the strange $D^*_{s0}$ and $D_{s1}$ shift considerably less and $\chi_{c0}, \chi_{c1}$ remain constant. A consistent pattern emerges across all four families: the hidden-charm states $\eta_c, J/\Psi, \chi_{c0}, \chi_c1$ fluctuate by no more than a few percent throughout the entire $\mu_B$ range at every temperature. As pure $c\bar{c}$ entities, they possess no light valence quark and experience the baryon chemical potential solely through the nearly frozen $\sigma_c$ -- a spectrum-level manifestation of the "Silver--Blaze" property \cite{Cohen2003,Braun2021}, which establishes charmonia as reliable probes of the surrounding matter. Each open-charm meson, conversely, responds precisely where the condensate dressing its light quark commences to dissolve: a $D$-type state transitions with $\sigma_x$, a $D_s$-type state with $\sigma_y$. Because this density dependence originates from the light condensates rather than from the absolute charm scale, it remains unaffected by the vacuum offset in the quarkonium masses noted above.

In accordance with the three-dimensional surfaces presented in Appendix~\ref{app:lowT} in Figs. \ref{Fig:7}--\ref{Fig:10}, three distinct layers of regularity can be identified here, which demonstrate consistency with the $\mu_B=0$ scenario. \begin{itemize}
\item First, regarding the channel classification, scalar and axial-vector mesons exhibit softening behavior, while pseudoscalar and vector mesons demonstrate hardening behavior, independent of flavor composition. This constitutes the spectral manifestation of chiral-partner convergence. 
\item Second, concerning flavor hierarchy, the magnitude of the shift decreases systematically from non-strange through strange to open-charm and finally to hidden-charm states, following the mass ordering of valence quarks and reflecting the dissolution of chiral condensates: the light $\sigma_x$ melts first, followed by the strange $\sigma_y$, and subsequently the charmed $\sigma_c$ \cite{SU34TempPRC}. 
\item Third, a distinct grouping pattern emerges is found so that hidden-charm states form an approximately flat ensemble, demonstrating insensitivity to variations in $\mu_B$ or temperature, whereas open-charm states exhibit a pronounced inflection point at the chiral phase transition. The $\eta--\eta'$ pair represents the sole exception to this pattern, attributable to the selective influence of the $U(1)_A$ anomaly.
\end{itemize}

In the scalar ($D_0^*, D^*_{s0}, \chi_{co}$), pseudoscalar ($D, D_s, \eta_c$), vector ($D^*, D_s^*,  J/\Psi$), and axial-vector ($D_1, D_{s1}, \chi_{c1}$) meson configurations, effective masses exhibit a stable plateau at low $T$ and low $\mu_B$. As these parameters are increased towards their extreme values, all meson states demonstrate a gradual reduction in mass, influenced by in-medium modifications, thermal fluctuations, and high-density polarization. As we already note above, all meson states seem to feel the flavor and quantum number hierarchies. The expected heavy-quark symmetry multiplets and the constituent quark content are reflected in the baseline mass scales, where open-charm states are generally found between $2100\text{ and }2600$~MeV. In contrast, strange-charmed states increase in energy proportionately, and hidden-charm/charmonium states ($J/\Psi, \chi_{co}, \chi_{c1}$) are located at elevated energy thresholds, specifically between $2400\text{ and }3980$~MeV. Sharp topological gradients and specific valleys located near the edges of the phase-space grids ($T$: from $100$ to $300$~MeV and $\mu_B$: from $0$ to $1400$~MeV) illustrate the strong association of charmed hadrons with melting chiral condensates and color-screening phenomena. And open charm mesons are sensitive to deconfinement and chiral phase transition boundary.

In this regard, we recall that the comprehensive examination of the four non-charmed meson configurations -- scalar ($a_0, K_0^*, \sigma, f_0$), pseudoscalar ($\pi, K, \eta, \eta'$), vector ($\rho, \omega, K^*, \phi$), and axial-vector ($a_1, f_1, K_1, f_{1s}$) -- provides significant insights into the behavior of light-quark hadrons under extreme thermodynamic conditions ($T=60\text{--}150\text{ MeV}$ and baryon chemical potentials $\mu_B$ reaching up to $1400\text{ MeV}$). The channel-specific responses can be classified as follows: 
\begin{itemize}
\item{Scalar Mesons:} The chiral partners $a_0$, $K_0^*$, and the broad $\sigma$ field exhibit deep minimum {\it ''valleys''} around $\mu_B\approx1000\text{ MeV}$, signifying critical softening and melting of the scalar condensate close to the chiral phase transition. The heavier $f_0$ meson state remains stable initially before decreasing steadily at high densities.
\item{Pseudoscalar Mesons:} As Goldstone bosons of chiral symmetry, the pion ($\pi$), kaon ($K$), and eta ($\eta$) masses remain flat at low densities before experiencing dramatic upward growth as $\mu_B$ exceeds $700\text{ MeV}$. Conversely, the flavor-singlet-mixed $\eta'$ drops sharply in mass, illustrating the strong impact of the partial restoration of the $U_A(1)$ axial anomaly.
\item{Vector Mesons:} The $\rho, \omega, K^*
$, and $\phi$ meson states all demonstrate smooth, uniform upward mass shifts as both temperature and baryon chemical potential increase, highlighting distinct medium polarization and vector-channel properties.
\item{Axial-Vector Mesons:} The $a_1$, $f_1$, and $K_1$ meson states all display relative stability at low $\mu_B$ followed by a sharp drop in effective mass near $\mu_B \approx 800\text{--}1100\text{ MeV}$, reflecting chiral partner convergence and medium-induced suppression. The isoscalar-strange $f_{1s}$ meson state exhibits an inverse trend, featuring a moderate upward inflection before dropping at large baryon chemical potentials.
\end{itemize}
		
The comparative analysis demonstrates that non-charmed meson spectra serve as highly sensitive indicators of chiral symmetry restoration and modifications within the medium. In contrast to the scalar, axial-vector, and $\eta'$ channels, which undergo significant mass reductions or critical softening linked to the dissolution of chiral condensates, pseudoscalar Goldstone modes and vector mesons display a notable increase in mass at elevated baryon densities. These contrasting behaviors provide a thorough foundation for comprehending hadron characteristics and chiral-axial dynamics.

Let us now discuss the overall properties of the numerical analyses which are expected to meet four consistency checks that are independent of parameters, along with the underlying mechanisms:
\begin{enumerate}
\item[(i)] \emph{"Silver--Blaze"}~\cite{Cohen2003,Braun2021}: At the onset, no observable is expected to depend on $\mu_B$. In fact, at a temperature of $T=100$~MeV, every state remains stable to within $<1\%$ up to $\mu_B\simeq 600$~MeV (excluding the soft $\sigma$). The corresponding baryon number density that can be derived as $n_B=-\partial\Omega/\partial\mu_B$ vanishes at $\mu_B=0$ and increases steadily thereafter, while the pure $c\bar c$ charmonia remains constant throughout.
\item[(ii)] \emph{Onset universality}~\cite{Parganlija2010,31}: States constructed from a shared condensate align together. Light and open-strange states correspond to $\sigma_x$, while the $s\bar s$ states correspond to $\sigma_y$. The $D$-type mesons are associated with $\sigma_x$, and the $D_s$-type mesons are linked to $\sigma_y$, whereas the charmonia remain unaffected. Consequently, the initial chemical potential is determined solely by the composition of particles.
\item[(iii)] \emph{Exact degeneracies}~\cite{Parganlija2010,31}: Under the isospin limit conditions where the values of $m_\rho=m_\omega=820$ and $m_{a_1}=m_{f_1}=1054$~MeV are likely held to computer machine precision for all $(\mu_B,T)$; the (axial-)vector states, which are expressed algebraically in $\sigma_x$, coincide precisely, while the $0^{+-}$ states undergo scattering fluctuations of a few tens of MeV due to the explicit constituent-quark loop~\cite{23}, leading to a broadening of around $\pm100$~MeV near the crossover at $T=150$~MeV.
\item[(iv)] \emph{Parity convergence}~\cite{DeTar1989,Rapp2000}: The vacuum splittings of $m_{a_1}-m_\rho \simeq 234$ and $m_{\sigma}-m_{\pi} \simeq 432$~MeV tend to zero as $\mu_B$ increases, reflecting the phenomenon of chiral restoration. Additionally, the $a_1$-$\rho$ gap remains less than the Weinberg value of $m_{a_1}=\sqrt{2}\times m_{\rho}$~\cite{Weinberg1967}.
\end{enumerate}

\section{Phase Boundary}
\label{sec:phaseboundary}
	
\begin{figure}
\centering
\includegraphics[width=1.0\linewidth]{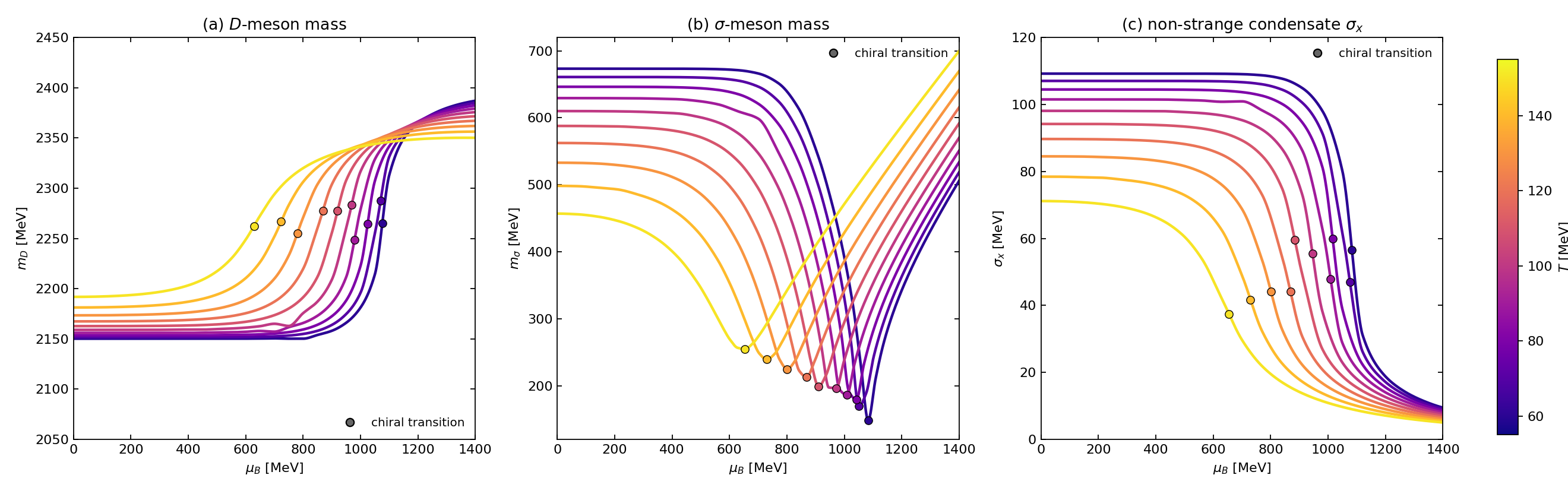}
\caption{The chiral transition deduced from three independent diagnostic probes in the SU(4) eLSM, is presented as a function of $\mu_B$ at ten temperature points between $T=60$ and $150$~MeV (symbols). Panel (a): The open-charm $D$-meson mass undergoes an elevation during the transition. Panel (b): The $\sigma$-meson mass possesses a soft-mode minimum that marks the transition location. Panel (c): The non-strange condensate $\sigma_x$ reaches its steepest descent at the transition. All three independent diagnostic methods converge on the identical critical baryon chemical potential $\mu_B^{c}(T)$.}
\label{Fig:Dmeson}
\end{figure}

%
Figure~\ref{Fig:Dmeson} presents the three independent diagnostic probes of the chiral transition in SU(4) eLSM model as functions of $\mu_B$ for different temperatures, ranging from $60$ to $150$~MeV: the open charm D-meson, the $\sigma-$ mass and the non-strange condensate $\sigma_x$. The open-charm sector does more than simply exhibit a mild density dependence; it serves as a precise indicator of the chiral transition itself. Figure~\ref{Fig:Dmeson} illustrates the $D$-meson masses plotted against $\mu_B$ for ten different temperatures, ranging from $T=60$ to $150$~MeV. Each curve displays a consistent two-stage profile, characterized by a flat plateau at low chemical potential, followed by an increase once $\mu_B$ reaches the chiral transition. The point at which this increase occurs shifts systematically to lower $\mu_B$ values as the temperature rises, moving from $\mu_B\simeq1076$~MeV at $T=60$~MeV to $\mu_B\simeq628$~MeV at $T=150$~MeV. The circle on each curve indicates the inflection point, defined as the baryon chemical potential at which the slope is steepest, i.e., pseudo-critical temperature or dissolution of hadron into parton, so that
\bea
\mu_B^{c}(T) &=& \arg\max_{\mu_B}\,\left|\frac{\partial m_D}{\partial\nu_B}\right|.
\eea
Given that the $D$-meson mass is ostensibly affected by the baryon chemical potential exclusively via the light condensate $\sigma_x$, which couples to its light quark constituent, this inflection point corresponds, subject to grid resolution constraints, with the dissociation of $\sigma_x$: consequently, the open-charm meson furnishes a unique, charm-identified marker of chiral-symmetry restoration.

\begin{figure}
\centering
\includegraphics[width=1.0\linewidth]{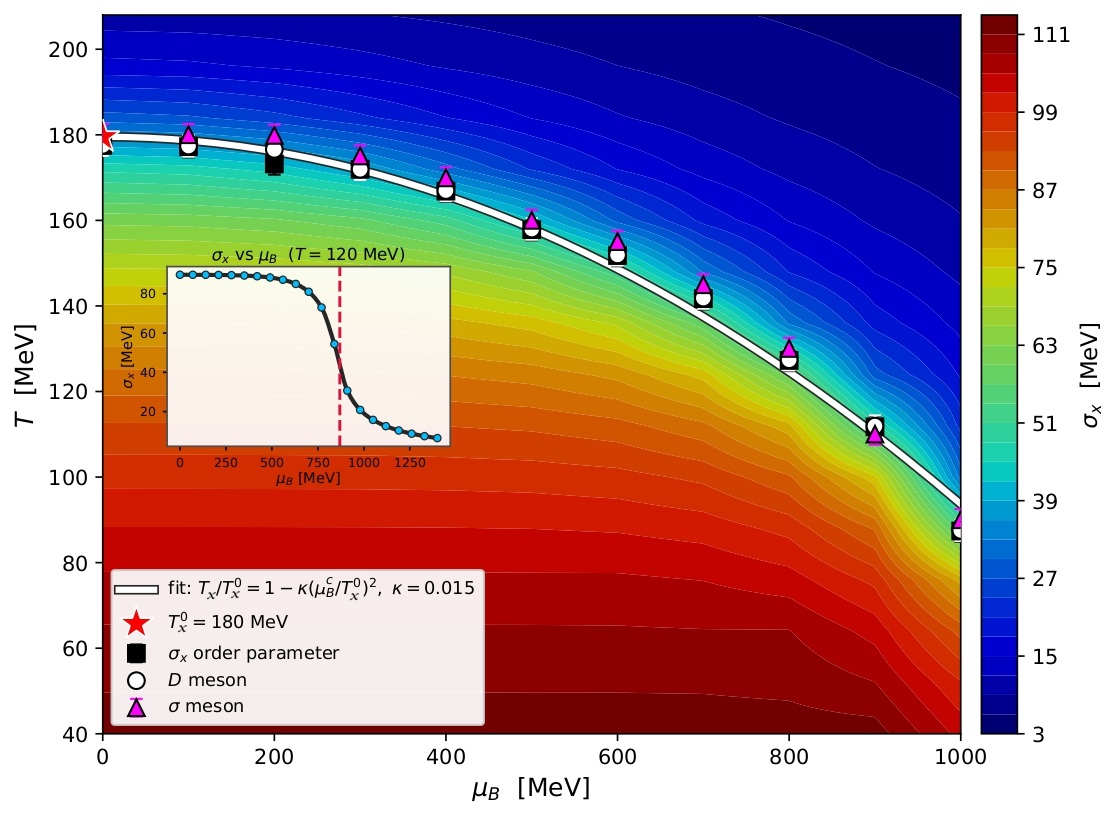}
\caption{Phase diagram of QCD in the $T-\mu_B$ plane, depicting the pseudo-critical boundary fitted by Eq.~\eqref{eq:curvature} with $T_c^0=180\text{ MeV}$ and $\kappa=0.015$. The background shading indicates the magnitude of the $\sigma_x$ order parameter. Symbols represent multi-probe extractions from $\sigma_x$ configurations (squares), $D$ mesons (circles), and $\sigma$ mesons (triangles). The inset displays the profile of the $\sigma_x$ order parameter versus $\mu_B$ at a fixed temperature $T=120$~MeV. }
\label{Fig:phaseboundary}
\end{figure}

%
Collecting the inflection points $\big(\mu_B^{c},T^{c}_{\chi}\big)$ maps out a line in the temperature-chemical-potential plane, shown in Fig.~\ref{Fig:phaseboundary}. This line is the chiral phase boundary as seen by the charmed sector. We parameterize it with the standard curvature expansion of the pseudo-critical line,
\begin{equation}
\frac{T^c_{\chi}(\mu_B)}{T_{\chi}^{0}} = 1 - \kappa \left(\frac{\mu_B^c}{T_{\chi}^{0}}\right)^{2} + \dots,
\label{eq:curvature}
\end{equation}
where $T_{\chi}^{0}$ is the transition temperature, at vanishing density and $\kappa$ the curvature of the boundary. A fit to the ten charm-determined points gives $T_{\chi}^{0}\simeq180$~MeV and $\kappa\simeq0.015$; the fit is anchored on the moderate-density points, where the mean-field description is reliable, and the curvature is stable ($\kappa\simeq0.015\pm0.002$) whether or not the two coldest, highest-density points are included. The value of $\kappa$ is of the same order as the continuum lattice-QCD determinations at small chemical potential, $\kappa\simeq0.012\pm0.016$~\cite{HotQCD2019,WB2020}; we stress that Eq.~\eqref{eq:curvature} is a small-$\mu_B$ expansion, whereas our points span $\mu_B/T\simeq 4$ to $15$, so the number quoted here is an \emph{effective} curvature over the density range explored rather than the Taylor coefficient measured on the lattice. Consistently with this, the two coldest points ($T=60,70$~MeV) fall slightly below the quadratic, indicating that the boundary steepens toward the high-density axis, as expected as the transition sharpens.
	
The phase boundary is mapped from $\mu_B=0$ up to approximately $1000$~MeV using all three probes of Fig. \ref{Fig:Dmeson}: the $\sigma_x$ order parameter (black squares),  $D$ meson ( circles) and $\sigma$ meson indicator (triangles). Throughout the examined region, these probes delineate a unified phase boundary. The transition parameterization, i.e., the dependence of the pseudo-critical temperature on the baryon chemical potential is characterized by Eq.~\eqref{eq:curvature}. The chiral limit baseline is defined at vanishing $\mu_B$ so that the baseline chiral crossover temperature is established at $T_{\chi}^0=180$~MeV. The curvature coefficient, namely the dimensionless curvature parameter is extracted as $\kappa=0.015$, reflecting a robustly stable pseudo-critical boundary under increasing baryon density. The order parameter evolution is depicted by the background color gradient, which represents the magnitude of the $\sigma_x$ order parameter (spanning from $3$~MeV to $111$~MeV, illustrating the gradual suppression of chiral condensate values from the hadronic phase into the deconfined regime. The embedded inset highlights the specific behavior of $\sigma_x$ as a function of $\mu_B$ evaluated at a constant temperature of $T=120$~MeV, capturing the sharp crossover drop across the chemical potential range. Finally, the multi-observable consistency is graphically represented by the numerical markers corresponding to individual $D$ meson and $\sigma$ meson tracking analyses, which align closely with the $\sigma_x$ contour, confirming universal scaling behavior across diverse hadronic probes.
	
A significant aspect of these findings is the remarkable consistency among the three independent microscopic probes employed to identify the chiral phase transition: the $\sigma_x$ chiral order parameter (black squares), the thermal behavior and mass changes of $D$ mesons (red circles), and the characteristics of the $\sigma$ meson (blue triangles). Throughout the entire examined region, these various probes delineate a unified phase boundary. At $\mu_B=0$, the pseudo-critical temperature attains a value of $T_{\chi}^0=180\text{ MeV}$, as marked by the red star on the temperature axis. The reduction in the pseudo-critical temperature as baryon chemical potential increases is described by a standard quadratic expansion fit to Eq.~\eqref{eq:curvature}. The derived curvature parameter is $\kappa=0.015$. This relatively low curvature value indicates a stable phase boundary at low to moderate baryon chemical potentials, which gradually curves downward in high-density regions. The strong correlation between the order parameter and the meson-based observables illustrates that effective chiral models that include heavy quark flavors ($\text{SU}(4)$ eLSM) offer a solid and internally coherent framework for investigating the QCD phase diagram restoration.

The essential message conveyed by Figs.~\ref{Fig:Dmeson} and~\ref{Fig:phaseboundary} is that the mass of the in-medium $D$ meson serves as a direct indicator of the chiral phase boundary at finite baryon density. While light mesons also experience the transition, they are integrated within the thermal medium, making their signals challenging to distinguish. In contrast, the hidden-charm states are entirely insensitive to $\mu_B$, as previously illustrated. Open charm occupies a unique intermediary position that enhances its utility: it is substantial, infrequent, and produced early enough to observe the dense phase of a collision, yet it possesses a single light quark whose condensate dissipates during the transition. The anticipated shift is significant: at $T=100$~MeV, the mass of the $D$ meson increases by approximately $200$~MeV across the transition, making it theoretically measurable. This scenario aligns perfectly with the high baryon density and moderate temperature conditions pursued by the CBM experiment at FAIR and by MPD at NICA, both of which include open-charm production in their research agendas \cite{Adzhymambetov2026EoS,Sinyukov2024iHKM,Xu2021QCDMatterReview}. The density dependence of the $D$-meson mass outlined here provides a tangible, charm-based observable to identify the chiral boundary in a domain that is not accessible to first-principles lattice calculations.
	
Two important caveats limit these conclusions. Firstly, the alignment of the $D$-meson inflection with the $\sigma_x$ transition arises from the light-quark dressing of the open-charm states, rather than being an independent phenomenon; this finding should be interpreted as open charm \textit{mapping} the light-sector transition, which provides an experimental avenue, rather than indicating a separate charm dynamics. Secondly, the $\sigma$ soft mode intensifies as the temperature decreases (its minimum drops from approximately $900$~MeV at $T=300$~MeV to around $170$~MeV at $T=70$~MeV) but does not disappear within the examined range; thus, we do not recognize a critical endpoint. We note that the phase boundary depicted in Fig.~\ref{Fig:phaseboundary} remains a crossover line throughout.

\section{Conclusions and outlook}
\label{sec:conclusion}
		
		
We have carried the SU(4) extended linear-sigma model from finite temperature to finite baryon chemical potential neglecting the vector interaction in quark sector and following to the standard single $\mu_B$ prescription. Considering two cases with $c=0$ and $c\neq 0$, we mapped the in-medium masses of the full meson spectrum (scalars, pseudoscalars, vectors and axial-vectors, across the light, strange and charmed sectors) as functions of $\mu_B$ for $T=100--300$~MeV. In the light and strange sectors, the results follow the expected course of chiral-symmetry restoration. The pseudoscalars rise while their scalar partners fall, the vectors and axial-vectors converge, and the $\sigma$ softens sharply at the transition. As the temperature grows, the onset $\mu_B$ moves inward, so the family of curves traces out the phase boundary. The strange states lag the non-strange ones because $\sigma_y$ outlasts $\sigma_x$. The charmed sector separates cleanly. The open-charm mesons acquire a mild but definite dependence on density. The $D$- and $D^\ast$-type states rise, the $D_0^\ast$- and $D_1$-type states fall, and each turns over at the transition of the lightest condensate it carries, so the strange charmed states move noticeably less than their non-strange partners. The hidden-charm states behave in the opposite way and stay essentially flat over the whole range of $\mu_B$ at every temperature. Carrying no light valence quark, they respond to the chemical potential only through the nearly frozen charm condensate. This is a spectrum-level realization of the Silver--Blaze property, and it singles the charmonia out as clean probes of the surrounding light matter. 

The spectra respect four parameter-independent checks: Silver--Blaze flatness, onset universality, the exact isospin degeneracies, and parity-partner convergence. Together these show that the trends are set by the condensates and not by the details of the fit. It is clearly seen that below onset, no observable depends on $\mu_B$, except the soft $\sigma$. Onset universality is confirmed by the fact thwt states built on a common condensate turn on together: light and open-strange states track light condensate, $s\bar{s}$ states track $\sigma_y$, D-type mesons track to condensates of the light flavour and charmonia stay inert. In the isospin limit the exact isospin degeneracies are observed. And the parity-partner covergence follows from the vacuum splitting $m_\rho - m_\omega \approx$ 234 MeV, $m_\sigma -m_\pi \approx $ 432 MeV shrink toward zero as $\mu_B$  grows.

We conclude that the integration of the multi-channel framework encompassing scalar, pseudoscalar, vector, and axial-vector charm sectors provides essential understanding of heavy-light and heavy-heavy quark system dynamics under extreme thermodynamic conditions:
\begin{itemize}
\item{Medium Modifications and Chiral/Thermal Sensitivity:} The detailed surface topographies demonstrate that charm meson masses undergo systematic medium modifications resulting from concurrent increases in temperature and quark chemical potential. Although heavy quarkonium states including $J/\Psi$ and $\eta_c$ maintain relative stability due to the heavy quark mass scale, open-charm states exhibit more pronounced environmental dependencies reflecting their interactions with the surrounding thermal and dense medium.
\item{Parameter Independence:} The persistent structural correspondence between the $c=11.24$ and $c=0$ surface layers across all examined meson states validates that qualitative physical phenomena, including critical slopes, contour trajectories, and threshold shifts, remain structurally invariant and independent of baseline energy offsets.
\item{Comprehensive Thermodynamic Blueprint:} This systematic comparison collectively establishes a rigorous multi-dimensional framework characterizing charm meson properties, delivering a fundamental theoretical foundation for comprehending heavy-flavor dynamics, in-medium bound-state stability, and QCD phase structures in nuclear matter.
\end{itemize}

The collective evaluation confirms that charmed meson properties are highly sensitive barometers for hot and dense QCD environments. The systematic survival of mass plateaus at low thermodynamic parameters, followed by smooth, coordinated downward shifts at high $T$ and large $\mu_B$ provides a robust quantitative framework for interpreting heavy-ion collision signatures, heavy-quark transport coefficients, and the approach to the deconfining phase transition. The distinction between density-sensitive open charm and near-inert hidden charm is particularly valuable: open charm maps the light-sector transition, while hidden charm serves as a clean reference gauge of the surrounding matter. It could be theoretically measurable shift, well matched to the high-baryon-density, moderate-temperature conditions pursued by CBM at FAIR and MPD at NICA.
		
The following directions follow naturally from what has been achieved in this work. The present equation of state was derived in the $g_\omega=g_\rho = 0$ limit, i.e. without vector mean-field contributions. Including them would give access to the full equation of state and to transport coefficients, extending the framework from spectroscopy toward hydrodynamics. The condensate trajectories computed here would serve as the background on which the vector fields are solved.The present spectrum contains only the ground state of each channel. Extending it to radial and orbital excitations would allow a direct check of Regge trajectories and of the in-medium behaviour of the string tension. The condensate trajectories and the phase boundary mapped here provide the medium background against which such an extension would be carried out.

\renewcommand{\thefigure}{A.\arabic{figure}} 
\setcounter{figure}{0} 
\setcounter{table}{0} 
\renewcommand{\thetable}{A\arabic{table}} 

\appendix

\section{Detailed $(T-\mu_B)$ Scan of Meson Masses}
\label{app:lowT}

\begin{figure}
\centering
\includegraphics[width=1.0\linewidth]{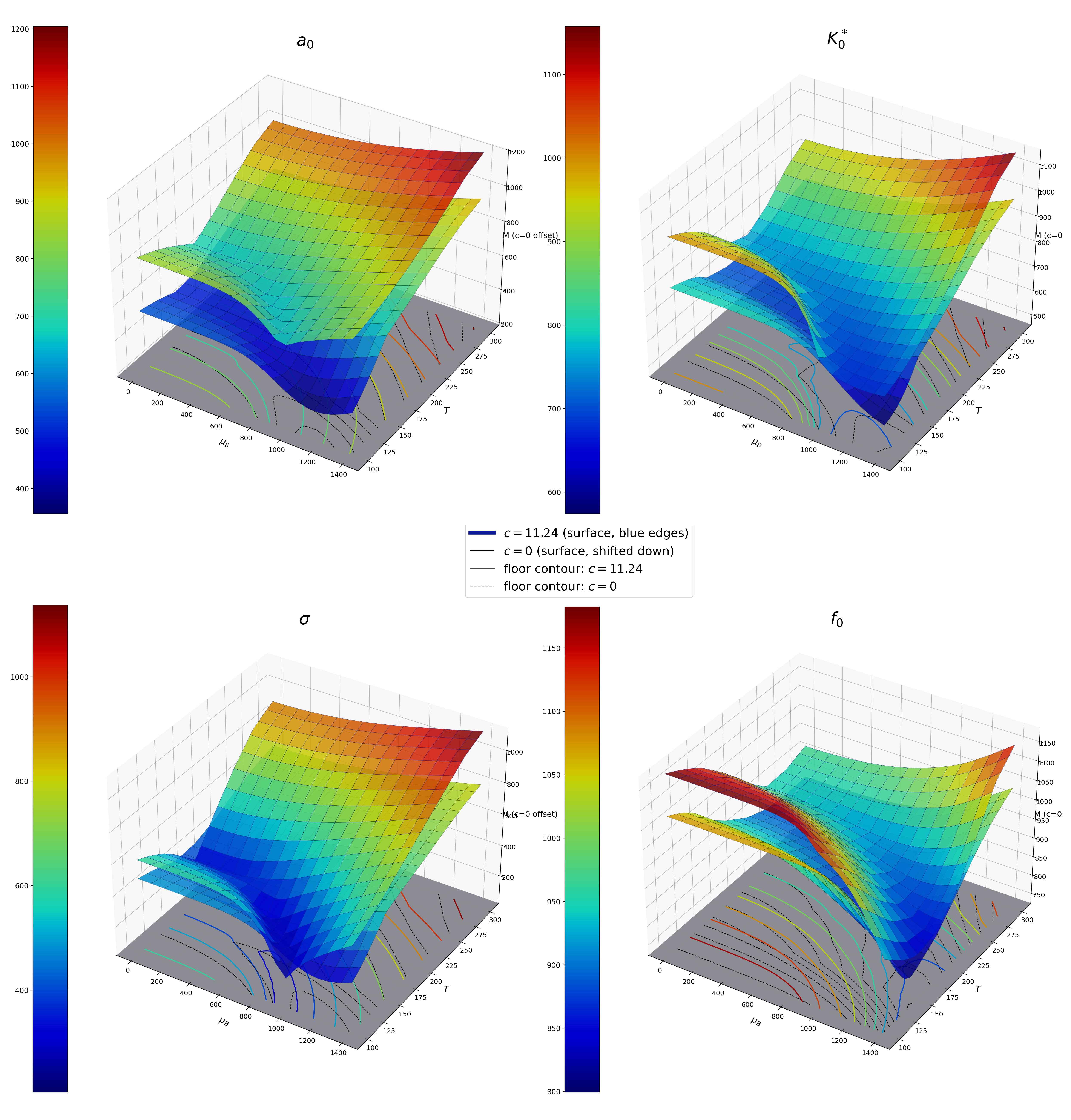}
\caption{A comprehensive three-dimensional depiction of the temperature ($T$) and baryon chemical potential ($\mu_B$) dependence of scalar meson masses distributed across four distinct channels: $a_0$ (top-left), $K_0^*$ (top-right), $\sigma$ (bottom-left), and $f_0$ (bottom-right). Each panel exhibits two configurations corresponding to distinct parameter values: the surface for $c=11.24$ (below surface) and the surface for $c=0$ (top surface). The projected floor contours delineate the parameter space behavior for both $c=11.24$ (solid lines) and $c=0$ (dashed lines) across the $T\text{--}\mu_B$ plane. Vertical color bars establish the magnitude scale of the respective scalar observables. }
\label{Fig:7}
\end{figure}
		
\begin{figure}
\centering
\includegraphics[width=1.0\linewidth]{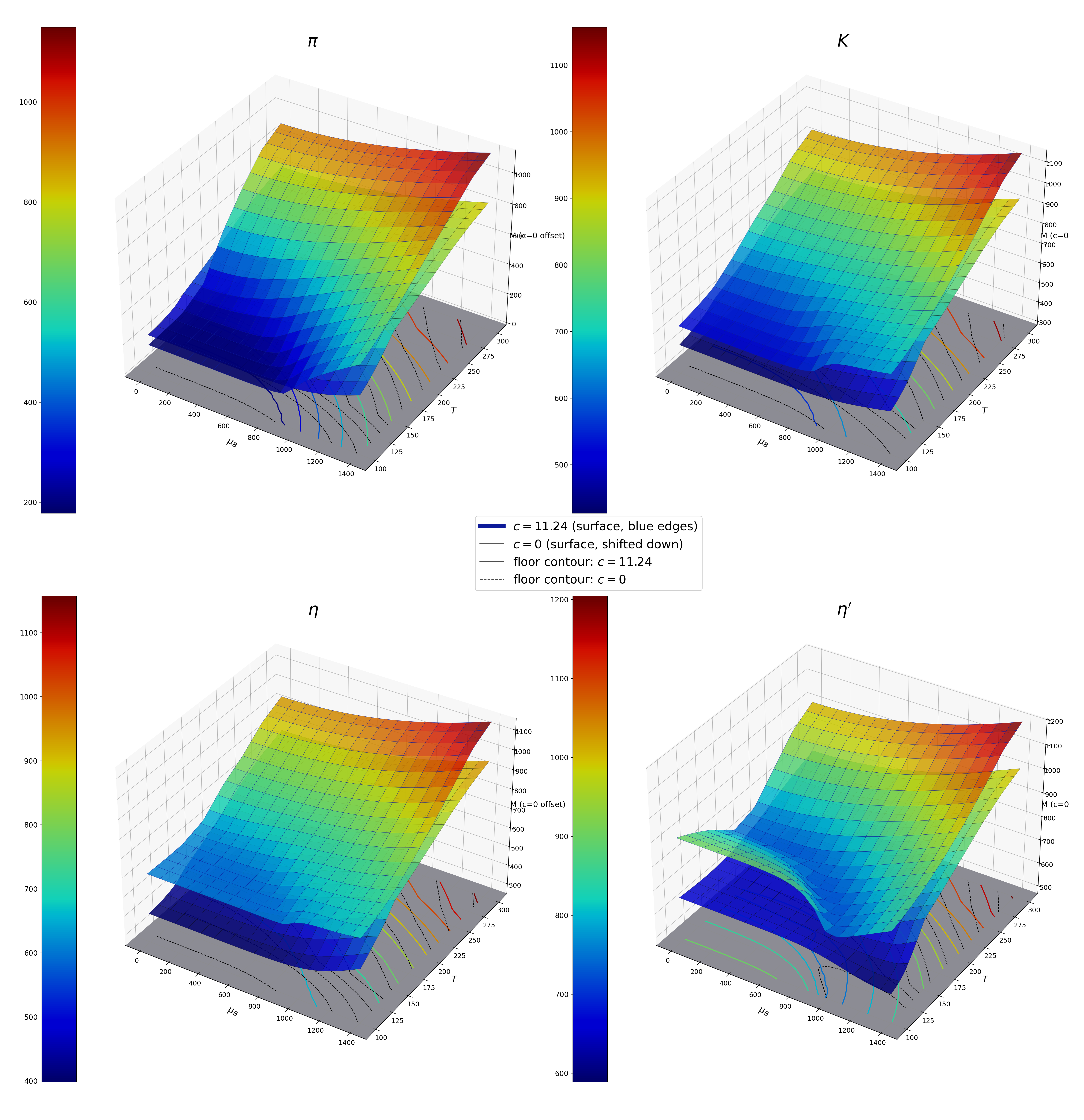}
\caption{The same as in Fig.~\ref{Fig:7} but here for the pseudoscalar meson masses ($\pi$, $K$, $\eta$, and $\eta'$).}
\label{Fig:8}
\end{figure}
				
\begin{figure}
\centering
\includegraphics[width=1.0\linewidth]{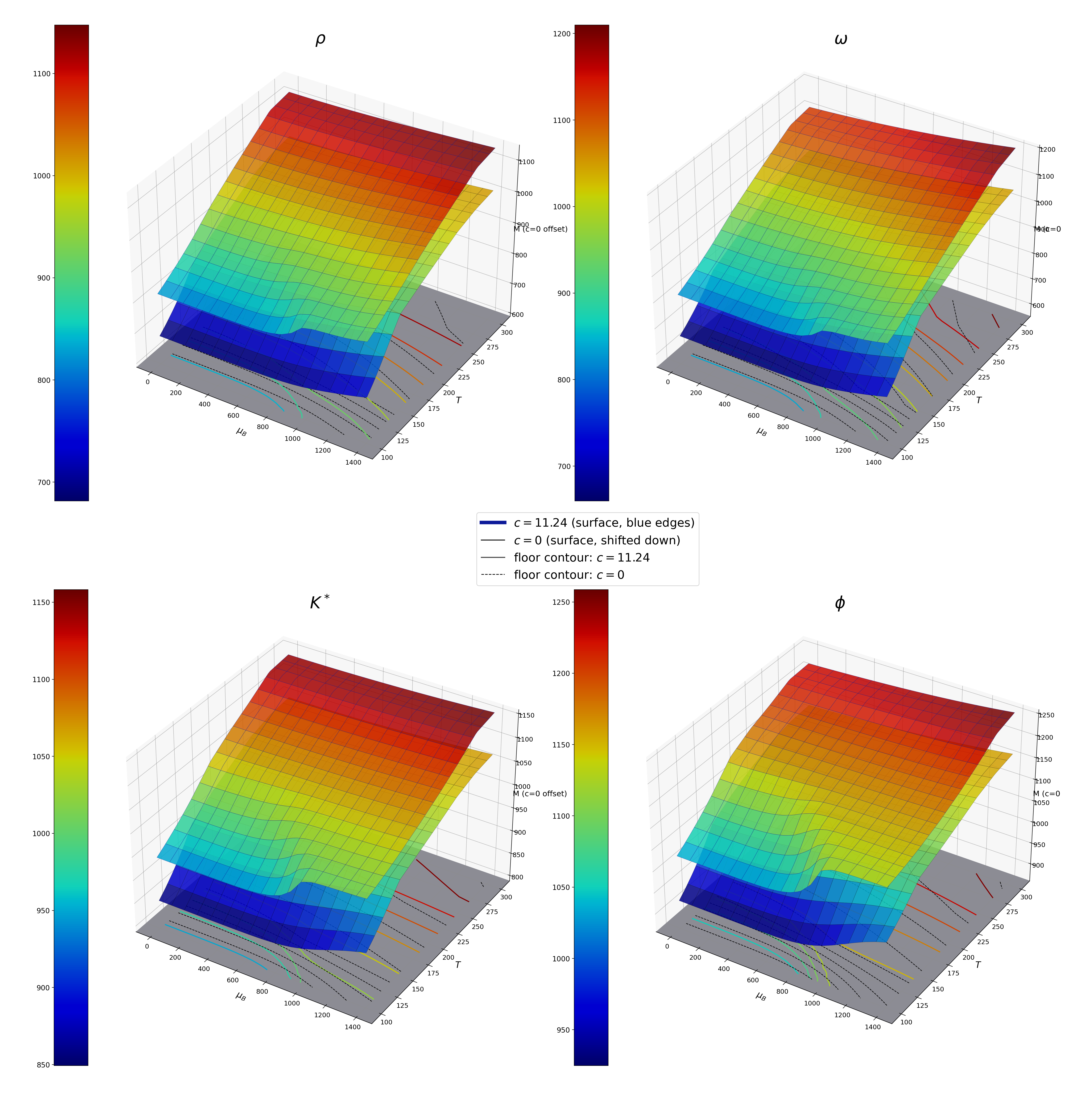}
\caption{The same as in Fig.~\ref{Fig:8} but here for vector meson effective masses ($\rho, \omega,  K^*, \phi$).}
\label{Fig:9}
\end{figure}
		
\begin{figure}
\centering
\includegraphics[width=1.0\linewidth]{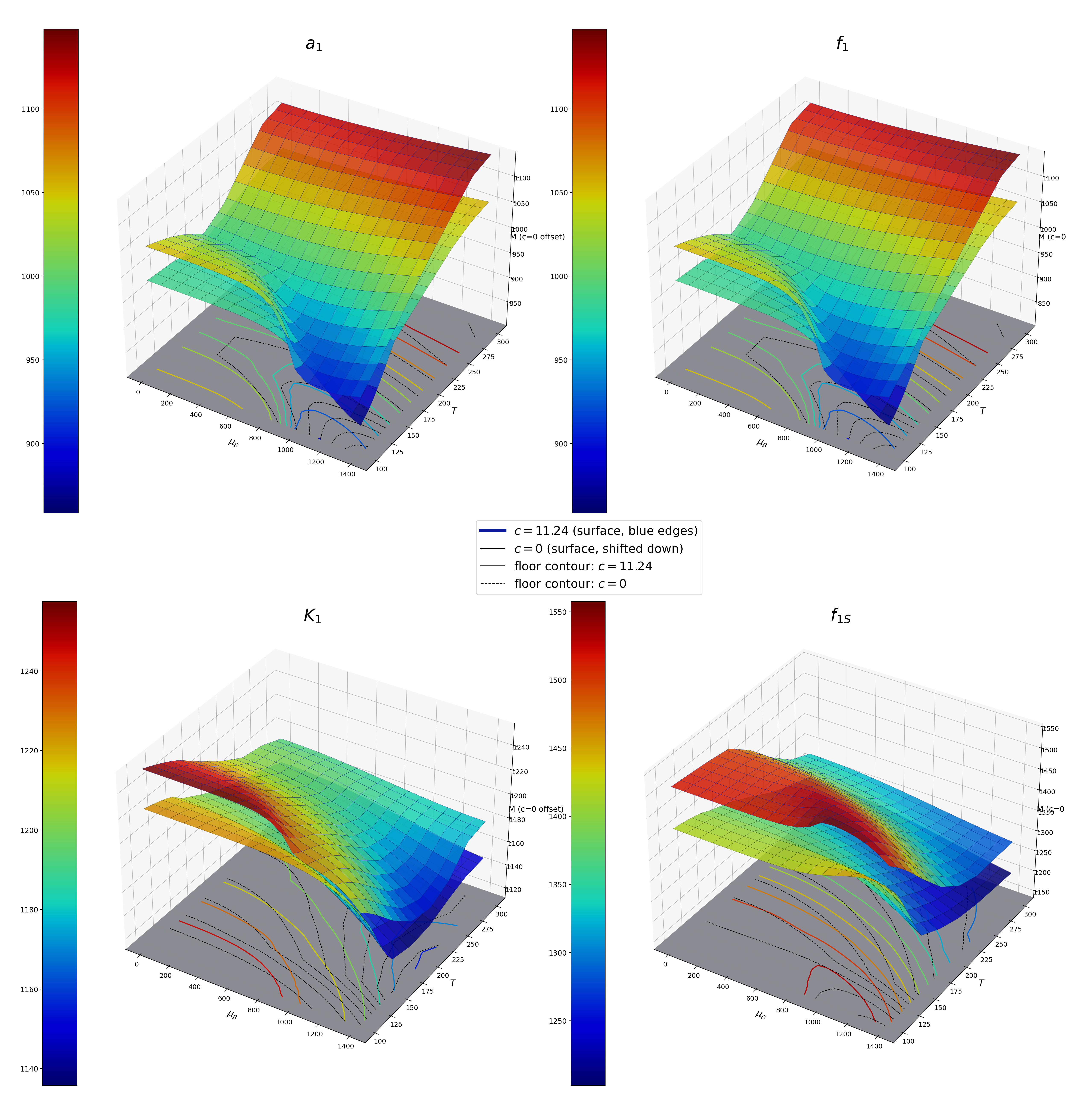}
\caption{The same as in Fig.~\ref{Fig:9} but here for axial-vector meson effective masses ($a_1$, $f_1$, $K_1$, and $f_{1s}$). }
\label{Fig:10}
\end{figure}

In the main text the in-medium modifications of messon masses are shown for $T=100--300$~MeV, the range relevant to the finite-density region of the phase diagram explored at RHIC-BES, FAIR and NICA \cite{Adzhymambetov2026EoS,Sinyukov2024iHKM,Xu2021QCDMatterReview} and where the mean-field Polyakov-loop description remains reliable \cite{Tawfik:2023egf,Tawfik:2021eeb,Tawfik:2019tkp}. At lower temperatures the chiral transition sharpens into a well-resolved bend, but it is pushed to $\mu_B\gtrsim1$~GeV, beyond the quantitative reach of the approximation. For the sake of completeness and to display the transition at its sharpest, we collect here the full meson spectrum as a function of the baryon chemical potential $\mu_B$ for ten temperatures $T=60--150$~MeV (color bar), organized by spin-parity family. The plateau, the bend at the transition and the recovery are all clearly resolved, and the onset moves to lower $\mu_B$ as $T$ increases, consistent with the chiral phase boundary of Fig.~\ref{Fig:phaseboundary}. Every state shows the behavior reported in Section~\ref{sec:discussion}: pseudoscalars and vectors harden, scalars and axial vectors soften, the $\sigma$ passes through its soft-mode minimum, the strange members lag the non-strange ones, and the hidden-charm states remain essentially inert.

We now evaluate the collective analysis of the four categories of charmed meson states: scalar, pseudoscalar, vector, and axial-vector (quarkonium), as illustrated in Figs.~\ref{Fig:7}--\ref{Fig:10}. This analysis offers a thorough understanding of heavy-quark dynamics within hot and dense hadronic matter. In all examined channels, which include light-charm, strange-charm, and charmonium systems, the thermodynamic response reveals consistent patterns influenced by chiral symmetry restoration and modifications due to the medium.	The three-dimensional surface plots presented in Figs.~\ref{Fig:7}--\ref{Fig:10} demonstrate the characteristics of diverse meson mass attributes as dependent variables of $T$ and $\mu_B$. All figures contain four subpanels representing particular meson channels belonging to distinct nonets, computed utilizing two separate configurations ($c=11.24$ and $c=0$) in conjunction with associated floor contours:
\begin{itemize}
\item{Scalar Meson Channels ($a_0, K_0^*, \sigma, f_0$):} Scalar field configurations -- specifically the chiral conjugate $\sigma$ and $a_0$ states -- demonstrate pronounced reductions aligned with chiral symmetry restoration phenomena. The dual-layer framework illustrates how adjustments to parameter $c$ modify the fundamental mass scale while preserving the essential structural characteristics.
\item{Pseudoscalar Meson Channels ($\pi, K, \eta, \eta'$):} The Goldstone and pseudo-Goldstone boson domain, encompassing the $\pi$ and $K$ modes, exhibits the anticipated low-energy robustness mandated by chiral symmetry requirements, whereas the flavor-singlet/octet combinations ($\eta, \eta'$) manifest abrupt curvature modifications reflecting both explicit and anomalous symmetry violation processes.
\item{Vector Meson Channels ($\rho, \omega, K^*, \phi$):} The vector nonet configurations demonstrate relatively modest fluctuations at reduced potentials, succeeded by downward adjustments at elevated temperatures and substantial quark chemical potentials, consistent with vector meson dominance phenomena and medium-induced modifications.
\item{Axial-Vector Meson Channels ($a_1, f_1, K_1, f_{1s}$):} These configurations characterize the temperature and density dependence of axial-vector meson resonances which seem to maintain considerable robustness at low temperatures and chemical potentials, succeeded by marked reductions or configurational transitions upon approaching the critical domain, with observable differentiation between strange and non-strange components.
\end{itemize}
		
An analysis of the multi-channel domain encompassing scalar, pseudoscalar, vector, and axial-vector sectors provides fundamental comprehension of QCD phenomena occurring at finite temperature and density:
\begin{itemize}
\item{Chiral Symmetry Restoration:} We note that all panels uniformly illustrate the softening of meson masses as the system approaches the critical phase boundary, serving as an effective indicator of chiral symmetry restoration.
\item{Parameter Robustness:} The simultaneous treatment of both $c=11.24$ and $c=0$ surfaces across all channels substantiates that the qualitative response to thermal and dense environments maintains structural integrity regardless of baseline offset variations.
\item{Sector-Dependent Sensitivity:} Light quark states devoid of strangeness exhibit steeper declines proximate to critical boundaries in contrast to their heavier strange-quark-bearing analogs (including $\phi$ and $f_{1s}$), resulting from explicit mass-generation processes.
\end{itemize}

The comprehensive topographical analysis encompassing all sixteen meson channels constructs an integrated thermodynamic representation, demonstrating that medium-induced modifications are substantially linked to the concurrent amplification of temperature and baryon chemical potential, thereby establishing a reliable theoretical framework for effective model applications in the context of dense nuclear matter.

\begin{figure}
\centering
\includegraphics[width=1.0\linewidth]{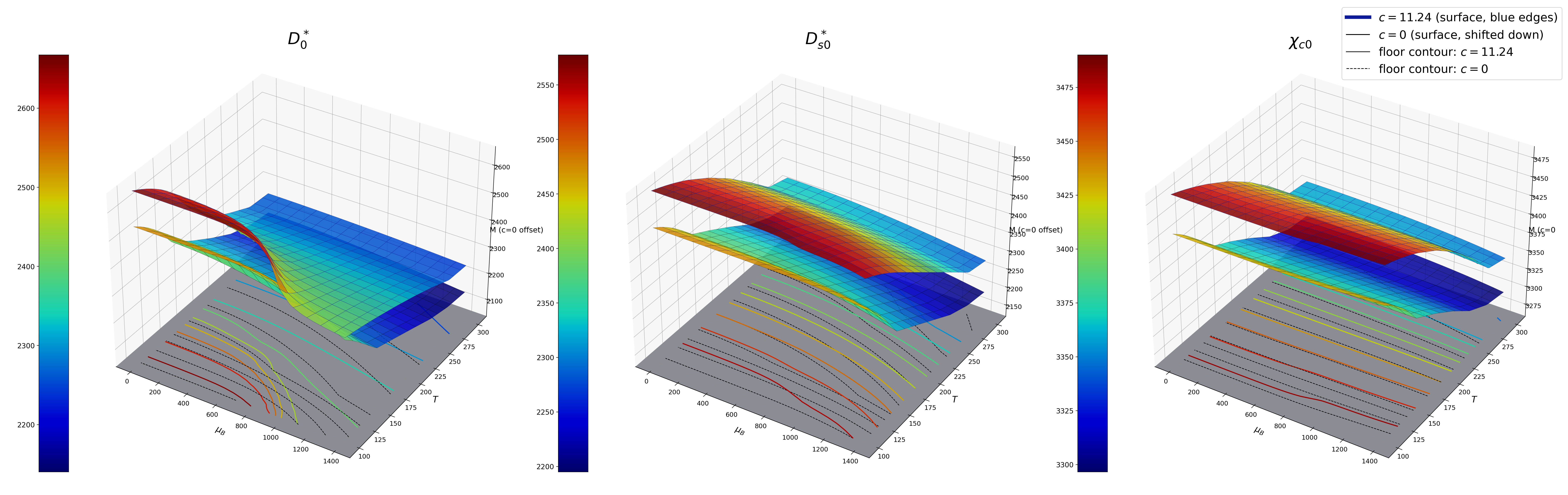} \\
\includegraphics[width=1.0\linewidth]{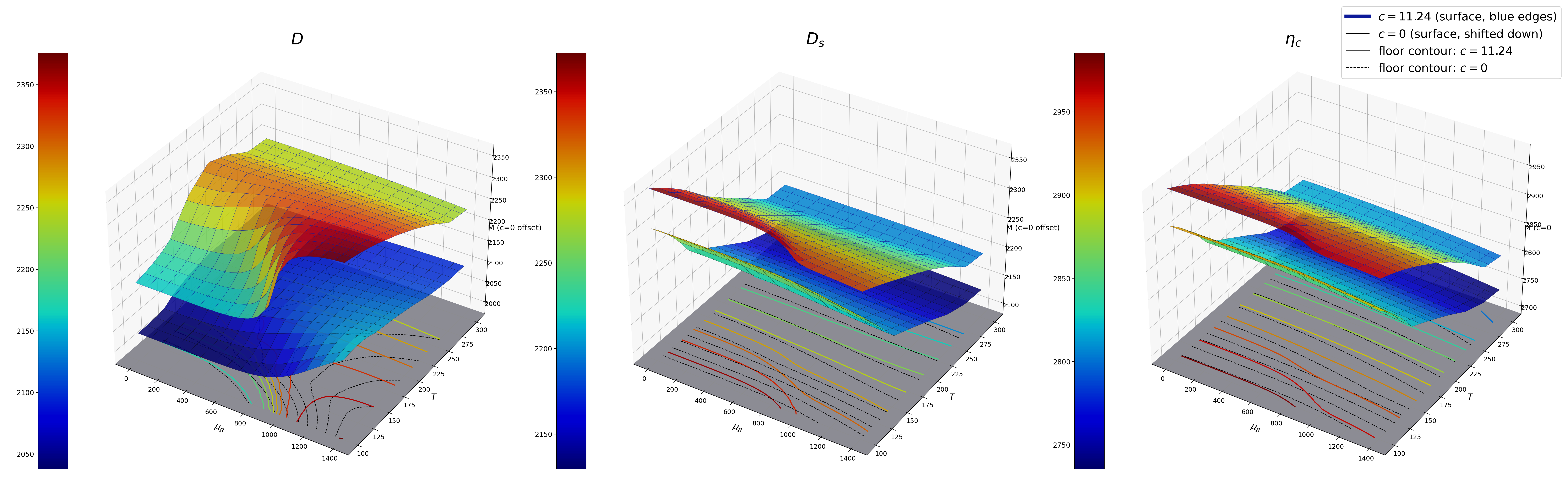} \\
\includegraphics[width=1.0\linewidth]{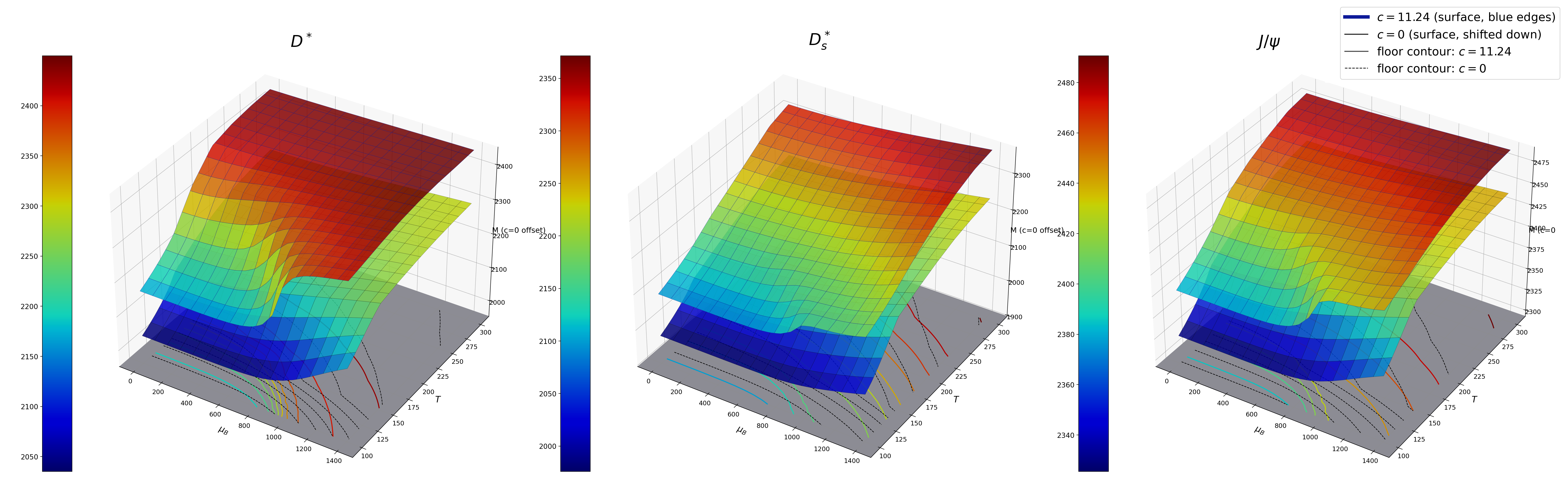} \\
\includegraphics[width=1.0\linewidth]{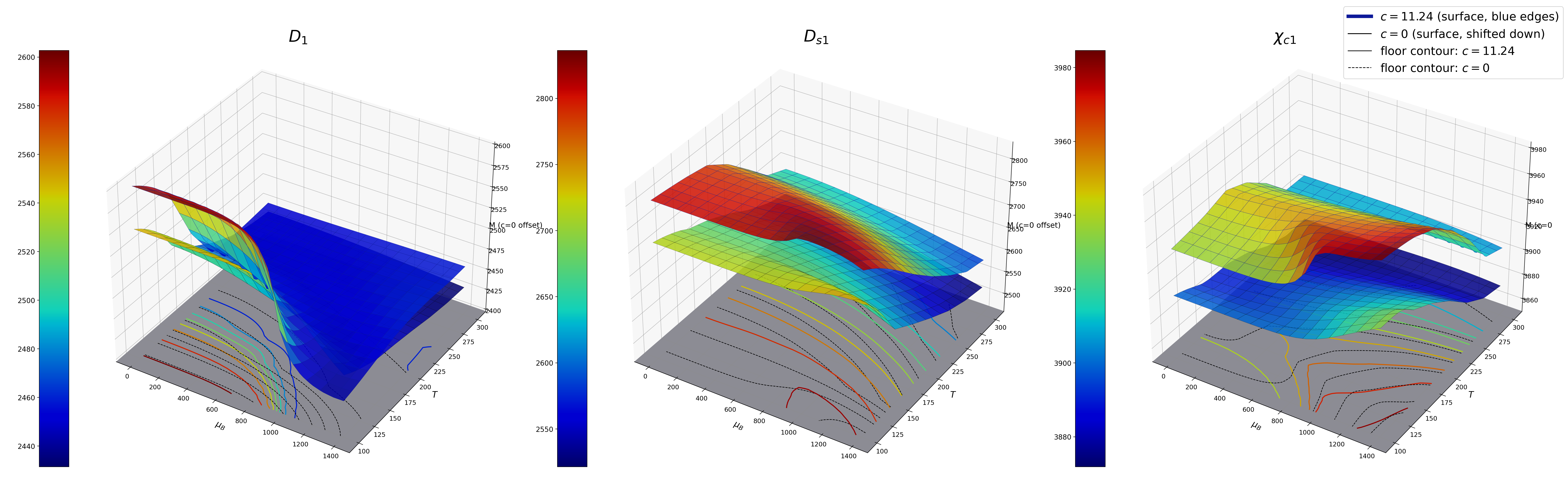} \\
\caption{A three-dimensional depiction of charm meson masses as functions of $\mu_B$ and $T$. Each panel showcases two separate configurations reflecting alternative parameter specifications: the upper surface representing $c=11.24$ and the lower surface representing $c=0$. Floor-level contour projections delineate the parameter space response for both $c=11.24$ (solid curves) and $c=0$ (dashed curves). }
\label{Fig:11}
\end{figure}

The three-dimensional plots constitute a thorough investigation of the thermal and density progression of charm meson mass parameters as parametric functions of $\mu_B$ and $T$. Each panel encompasses three separate subpanels denoting particular charm meson states, examined under two distinct parameter frameworks:  $c=11.24$ and  $c=0$. The vertical color indicators quantify the distribution and directional gradients of the pertinent physical quantities throughout the parameter space.
\begin{itemize}
\item Scalar charm meson sector ($D_0^*$, $D_{s0}^*$, and $\chi_{c0}$) manifests characteristic mass plateaus at low densities that progressively bend and decline within particular critical zones of the thermodynamic parameter space. The parallel positioning of the $c=11.24$ and $c=0$ surfaces substantiates the stability of the underlying effective model formulation with respect to parameter variations.
\item Pseudoscalar charm meson sector encompassing the open-charm $D$ and $D_s$ mesons in conjunction with the hidden-charm charmonium ground state $\eta_c$, exhibits smooth variations at low-to-moderate densities, transitioning into more pronounced slopes as temperature and quark chemical potential both increase. The contour lines projected onto the floor grid illuminate distinct thermal gradients and threshold behaviors intrinsic to the flavor content of each state.
\item Vector charm meson sector and heavy quarkonium channels, notably encompassing the vector $D^*$ and $D_s^*$ states in addition to the celebrated $J/\Psi$ charmonium meson, present relatively stable behavior at low potentials followed by gradual or moderately sharp medium-induced modifications at elevated temperatures and high quark chemical potentials, underscoring the resilience of heavy quark bound states in dense matter.
\item Axial-vector charm meson sector, featuring the $D_1$, $D_{s1}$, and $\chi_{c1}$ states, demonstrates considerable stability at lower temperatures and chemical potentials, followed by pronounced structural shifts, declines, or curvatures as the system approaches critical boundary regions. The dual-layer representation distinctly illustrates how the baseline offset parameter $c$ uniformly shifts the absolute energy scale without affecting the qualitative topographical trends across the ($\mu_B--T$) plane.
\end{itemize}

\bibliographystyle{unsrtnat}
\bibliography{2026-23-23-NourhaneSecondPaper}

\end{document}